 \documentclass[final,1p,times]{elsarticle}

\usepackage{graphicx}
\usepackage{graphicx}
\usepackage{subcaption}

\usepackage{hyperref}

\usepackage[utf8]{inputenc}
\usepackage[table]{xcolor}

\usepackage{footnote}
\makesavenoteenv{tabular}

\usepackage{booktabs,caption}
\usepackage[flushleft]{threeparttable}

\usepackage[linesnumbered,lined,boxed,commentsnumbered]{algorithm2e}

\usepackage[utf8]{inputenc}
\usepackage{xcolor}

\usepackage{alphalph}

\usepackage{comment}
\usepackage{makecell}

\usepackage{float}

\usepackage{amsmath}
\usepackage{xspace} 

\usepackage[misc,geometry]{ifsym} 

\usepackage{tikz}
\def\checkmark{\tikz\fill[scale=0.4](0,.35) -- (.25,0) -- (1,.7) -- (.25,.15) -- cycle;} 

\usepackage{comment}

\usepackage{hyperref}

\journal{Data \& Knowledge Engineering}

\begin{document}

\begin{frontmatter}

\title{FaNDS: Fake News Detection System Using Energy Flow}


\author[First]{Jiawei Xu\corref{corresponding}}
\cortext[corresponding]{Corresponding author.}
\ead{jix20@pitt.edu}

\author[First]{Vladimir Zadorozhny}
\ead{vladimir@sis.pitt.edu}

\author[First]{Danchen Zhang}
\ead{daz45@pitt.edu}

\author[Third]{John Grant}
\ead{grant@cs.umd.edu}

\address[First]{Department of Informatics and Networked Systems, School of Computing and Information, University of Pittsburgh, PA, USA}

\address[Third]{Department of Computer Science and UMIACS, University of Maryland, MD, USA}

\begin{abstract}
Recently, the term "fake news" has been broadly and extensively utilized for disinformation, misinformation, hoaxes, propaganda, satire, rumors, click-bait, and junk news. 
It has become a serious problem around the world.
We present a new system, FaNDS, that detects fake news efficiently.
The system is based on several concepts used in some previous works but in a different context.
There are two main concepts: an Inconsistency Graph and Energy Flow.
The Inconsistency Graph contains news items as nodes and inconsistent opinions between
them for edges.
Energy Flow assigns each node an initial energy and then some energy is propagated
along the edges until the energy distribution on all nodes converges.
To illustrate FaNDS we use the original data from the Fake News Challenge (FNC-1).
First, the data has to be reconstructed in order to generate the Inconsistency Graph.
The graph contains various subgraphs with well-defined shapes that represent 
different types of connections between the news items.
Then the Energy Flow method is applied.
The nodes with high energy are the candidates for being fake news.
In our experiments, all these were indeed fake news as we checked each using several
reliable web sites.
We compared FaNDS to several other fake news detection methods and found it to be more sensitive in discovering fake news items.
\end{abstract}


\end{frontmatter}


\section{Introduction}
\label{sec:intro}

Recently, the term "fake news" has been broadly and extensively utilized for disinformation, misinformation, hoaxes, propaganda, satire, rumors, click-bait, and junk news\cite{pierri2019false}. 
While people generally agree that all these terms indicate deceptive information, some researchers argue that a precise and agreed definition of fake news is still missing\cite{pierri2019false}. Some other researchers define fake news in a more restrictive manner as news articles that are intentionally composed to mislead and misinform readers and whose falsity is verifiable as false using other sources  \cite{allcott2017social,shu2019beyond}. As concluded in \cite{bondielli2019survey}, three major factors of fake news can be identified: (i) its form as a news article; (ii) its misleading intent, that could be malicious or satirical; and (iii) its verifiable content as partially or completely fake.

Compared to fake news, rumors have been more extensively studied as false information on the internet in recent research. Rumors relate to information that has not been authorized by official sources yet and is spread generally by users on social networks \cite{bondielli2019survey}. Various research studies have attempted to classify rumors with regard to the type, scope, along with its characteristics. Thus, there are various definitions for rumors. For example, in the research of \cite{difonzo2007rumor}, rumors are labeled as "unverified and instrumentally relevant information statements in circulation". And more specifically in \cite{zubiaga2015towards}, a rumor is defined as a “circulating story of questionable veracity, which is apparently credible but hard to verify, and produces sufficient skepticism and/or anxiety”. 

The plan of this paper is as follows.
In the rest of this section we comment on the history of fake news and how more
recently fake news has become ubiquitous on social media.
We also present some statistics about the number of research papers since 2010 
both on rumor detection and fake news detection.
Section~\ref{sec:relwork} gives a review of related work both on fake news 
detection and flow models.
Section~\ref{sec:background} covers the key background concepts: Inconsistency Graph (IG)
and Energy Flow (EF).
Then in Section~\ref{sec:method} we explain how to create the Inconsistency Graph 
and explains how we use the Energy Flow method.
Section~\ref{sec:experiments} describes the experimental method, our results, an
analysis of FaNDS and a comparison with other methods showing its higher sensitivity.
The paper ends in Section~\ref{sec:conclusion} with a summary and future work.

Both fake news and rumors are not really new items of the Internet era. For example, in April 1835, the New York Sun newspaper published a series of fake news articles stating that a famous astronomer found life on the moon \cite{GreatMoonhoax}. It is known as the Great Moon Hoax. Even research on rumors started long ago: its study can be traced back to the end of WWII\cite{Allport1946501,Allport1947240}. However, with the development of the internet and social media, fake news and rumors have increased exponentially in our world. 

The Internet in general, and Social Media in particular, with a massive number of users, has become the main source of news and information to large segments of the population. 
For this reason, the role of television and newspapers as information channels has greatly diminished. Social media like Facebook, weChat, TikTok, and Twitter have a huge number of active users worldwide. As shown in Statista.com, the number of active user accounts in October 2019 for the four social media above is 2414, 1133, 500, and 330 in millions, respectively\footnote{\href{https://www.statista.com/statistics/272014/global-social-networks-ranked-by-number-of-users/}{https://www.statista.com/statistics/272014/global-social-networks-ranked-by-number-of-users/}}. 
While the original purpose of social media was to find and maintain relationships with friends, it evolved into a news gathering source as well.
In fact, according to
Zubiaga et al. \cite{zubiaga2018detection}, social media have grown to be a crucial publishing tool for journalists \cite{diakopoulos2012finding,tolmie2017supporting} as well as the main way for readers to obtain the most up-to-date news \cite{hermida2010twittering}.

Social media has proved to be very helpful in some ways, especially during crisis situations, since it can spread breaking news much faster and more efficiently than the traditional media, such as TV and newspapers \cite{vieweg2010microblogged}.
However, there is a significant disadvantage as well: low (or no) barriers for posting
results in a lack of control and fact-checking. 
This way, the spread of news and information through social media often results in low quality, unverified, and even fake news\cite{allcott2017social,zubiaga2015towards}. 
As making content online is so simple and fast, entering the social media industry has a much lower barrier than the traditional media industry \cite{allcott2017social}. This has led to a decline of traditional journalistic standards as well as to a lack of third-party fact-checking \cite{allcott2017social,pierri2019false}.

There may be various motivations for spreading fake news. First, some fake news articles can draw considerable advertising and marketing profits for their providers \cite{allcott2017social}. Second, the providers of fake news may aim to influence public thoughts and opinions on certain subjects for political purposes \cite{allcott2017social}. In addition, the existence and growing number of deceptive agents, such as robots/bots, crawlers, and trolls, has been considered as another major source for spreading fake news and rumors \cite{shao2018spread,kumar2018false}.

The negative influence and effect of fake news and rumors has substantially increased in recent years, as social media facilitates its propagation to a large number of people very fast, significantly influencing public opinion and the understanding of certain events \cite{zubiaga2018detection}. For example, in the 2016 US presidential election, fake news concerning Hillary Clinton may have had a substantial impact on the result \cite{allcott2017social,pierri2019false}. Similarly, recent studies have shown that fake news also influenced the UK Brexit referendum \cite{howard2016bots} in 2016, and the French presidential election in 2017 \cite{ferrara2017disinformation}.

Fake news may also have a financial impact. In 2013, a fake tweet mentioned that Barack Obama had been injured in an explosion at the White House. According to the Financial Times, that one tweet caused a drop in the stock market with the S\&P 500 declining 0.9\% — enough to wipe out \$130 billion in stock value in a matter of seconds \cite{matthews2013does}.
Fake news can have serious side effects. 'Pizzagate' was a famous fake news item in the 2016 US elections to defame Clinton. It also resulted in gun violence in a restaurant as one of the people fooled by the hoax decided to investigate it. Eventually, the gunman was sentenced to 4 years in prison.\cite{shu2018fakenewsnet,hauck2017pizzagate}
Like fake news, spreading rumors can also lead to serious damage. For example, a rumor about a shooting near a school and the kidnapping of children in Veracruz caused car crashes as parents were rushing to pick up their kids \cite{Ma20163818}.


    \begin{figure*}
        \centering
        \begin{subfigure}[b]{1\textwidth}
            \centering
\includegraphics[width=\linewidth]{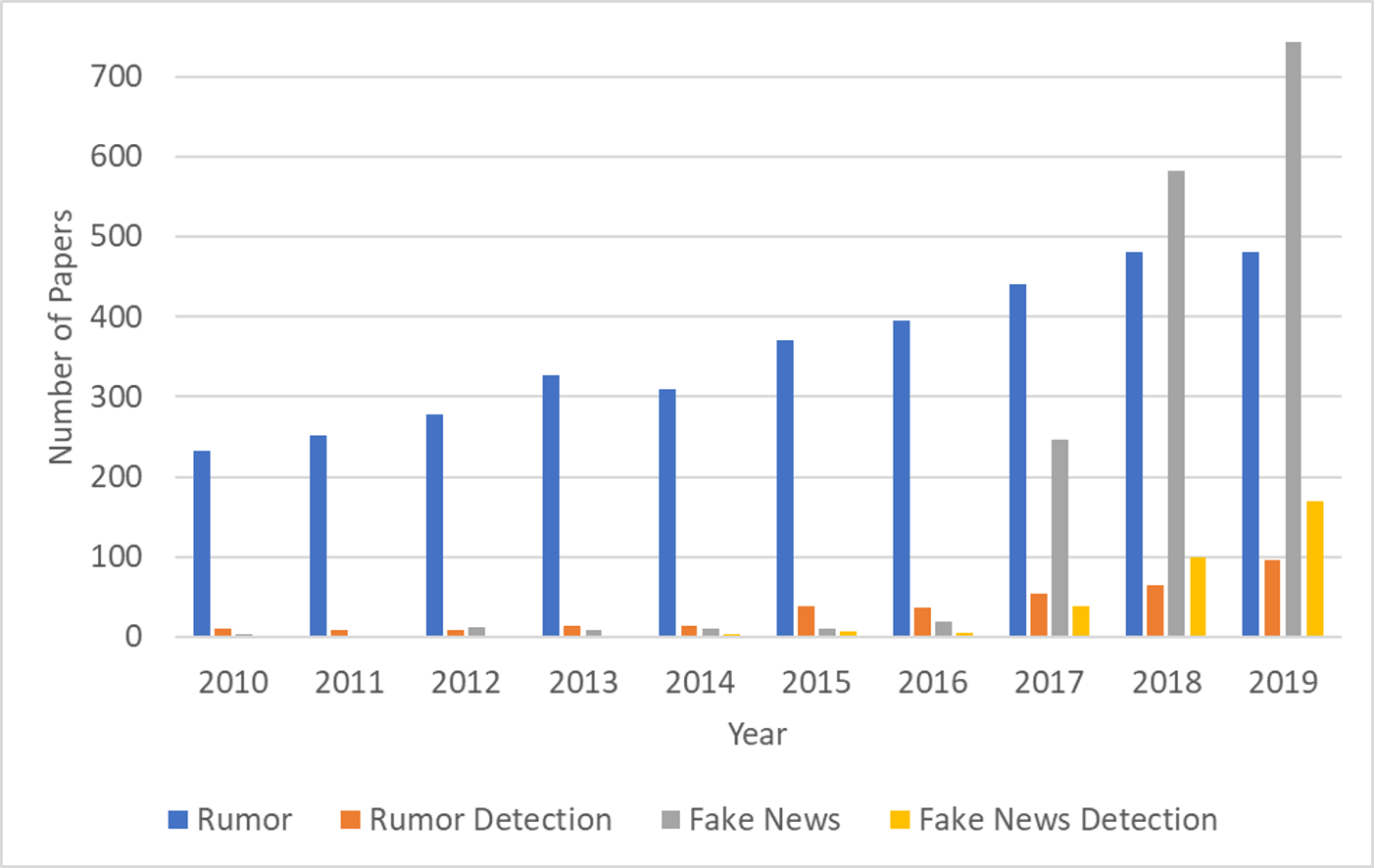}
        \end{subfigure}
        \par
        \caption{The number of research papers related to Rumor, Rumor Detection, Fake News, and Fake News Detection between 2010 and 2019}
	\label{fig:fake-news-analysis}
    \end{figure*}	 

In the past few years, the interest of researchers about fake news, rumors, and the technology of their detection has grown substantially. Figure~\ref{fig:fake-news-analysis} shows the trend of the number of research papers related to Rumor, Rumor Detection, Fake News, and Fake News Detection in the past 10 years (2010-2019), indexed through the Scopus Database \footnote{www.scopus.com}. Clearly, interest about rumors (blue) has been growing continuously, while fake news (gray) started to draw significant attention after 2016, probably due to its influence on the 2016 US presidential election. The trend of interest in rumor detection (orange) and fake news detection (yellow) is similar, but with fewer papers. Compared to rumor detection, while the first paper about fake news detection appeared much later (2013 vs 1992), the interest in fake news detection is growing much faster.

\section{Related Work}
\label{sec:relwork}
In this section, we first review research on techniques of fake news detection and then go through the major flow models and their applications.

\subsection{Fake News Detection}
Several recent survey papers encompass the wide range of research devoted to fake news  including \cite{kumar2018false}, \cite{shu2017fake}, \cite{conroy2015automatic}, \cite{chen2015misleading} and \cite{bondielli2019survey}. The most important problem in this area is to detect fake news automatically or to find the ones most worthy to be further checked.
As mentioned in the previous section, there are different types of fake news and there is also a close connection with rumors. So fake news detection techniques have a substantial overlap with the detection of rumors, fake opinion, fake accounts, hoaxes, and frauds. For that reason we include some algorithms from papers about those that can also be used for fake news detection.

Fake news detection uses primarily three kinds of information: (1) the content of news articles, including at the word, syntactic, and semantic levels, (2) news propagation by users on social networks, including user profiles, news profiles, spreading data, etc, and (3) network structure extracted from news articles and social media. In most cases, detection is implemented by a classification model on different kinds of features.

Word level and syntactic level features of news articles are found to be the most effective in many papers, such as the models in  \cite{rubin2015truth} and \cite{wang2017liar}. Both word and syntactic information are essential. The increasing popularity of neural networks in natural language processing (NLP) has led to the extraction and use of semantic features in fake news detection, such as the models in \cite{hassan2015detecting},\cite{potthast2017stylometric}, \cite{perez2017automatic}, \cite{ajao2018fake}, \cite{kochkina2018all}, \cite{song2019ced}, and \cite{zubiaga2018discourse}.

From another aspect, many researchers explore information from social networks where the news is spread and to the people in the network. They focus on news profile features, such as the number of likes and propagation times, and user profile features, such as the number of posts, registration age, and the number of followers. Many studies have found that systems cannot detect fake news accurately if they use only social network features, so they are usually used together with the content features, such as models in \cite{castillo2011information}, \cite{chu2010tweeting}, \cite{qazvinian2011rumor}, \cite{kwon2013prominent}, \cite{ma2015detect}, \cite{kumar2016disinformation}, \cite{liu2019towards}, and \cite{li2019rumor}.

Multiple network structures can be obtained from this area, such as user-follow-user networks, news-agree/conflict-news networks, and user-spread-news networks \cite{gupta2012evaluating}, \cite{jin2016news}, \cite{ruchansky2017csi}, \cite{tacchini2017some}, \cite{della2018automatic}, and \cite{guacho2018semi}.
There is also a smaller number of works that focus on news fact checking, where the reference facts are in a preexisting knowledge base such as DBpedia \cite{wu2014toward}, \cite{ciampaglia2015computational}, and \cite{shi2016fact}.

Our system, FaNDS, builds a graph from data with the news items as the nodes, and news stance conflicts as the edges. We run an energy flow model on the graph, and the ones with the highest energy are the news that should be checked with the highest priority. From the above mentioned works, we selected six methods and compare them with ours in Table \ref{table:compare_related_work}.

Both FaNDS and the model in \cite{hassan2015detecting} focus on finding the most check worthy news items. The latter extracts linguistic and semantic features from the claim and learns a classification model with well labeled training data. Jin et al. extracted a three-layer (messages, subevents, events) hierarchical network from Microblog, and ran a flow model on this three-layer network to evaluate the credibility of the news in \cite{jin2014news}. Then later, this group worked on Twitter data, and extracted stance conflicts among messages to enhance their graph structure. This resulted in better performance  \cite{jin2016news}. Gupta et al. proposed BasicCA and EventOptVA in \cite{gupta2012evaluating}. They extracted a graph with users and news from Twitter, gave initial credibility scores to user and news, and let the score propagate iteratively to get the final credibility evaluation for the news. Their model did not utilize the twitter content conflict information. Our model is a good addition to the current fake news detection efforts.




\begin{table}[]
\centering
\caption{Comparison of most related Fake news detection studies}

\begin{tabular}{|l|c|c|c|c|c|c|c|}
\hline

\multicolumn{1}{|c|}{Models} & \begin{tabular}[c]{@{}l@{}}FaNDS\end{tabular} & \begin{tabular}[c]{@{}l@{}}Hassan\\et al.\\\cite{hassan2015detecting}\end{tabular}  & \begin{tabular}[c]{@{}l@{}}Jin\\et al.\\\cite{jin2014news}\end{tabular} & \begin{tabular}[c]{@{}l@{}}CPCV\\\cite{jin2016news} \end{tabular} & \begin{tabular}[c]{@{}l@{}}BasicCA\\\cite{gupta2012evaluating}\end{tabular} & \begin{tabular}[c]{@{}l@{}}Event\\OptVA\\\cite{gupta2012evaluating}\end{tabular} & \begin{tabular}[c]{@{}l@{}}CCRF\\\cite{mukherjee2015leveraging}\end{tabular}  \\ \hline

\begin{tabular}[c]{@{}l@{}}Find check\\worthy news.\end{tabular} & \checkmark & \checkmark &   &   &   &   &   \\ \hline
\begin{tabular}[c]{@{}l@{}}Use flow models.\end{tabular} & \checkmark &   & \checkmark & \checkmark & \checkmark & \checkmark &   \\ \hline
\begin{tabular}[c]{@{}l@{}}Organize data in\\graph structure.\end{tabular} & \checkmark &   & \checkmark & \checkmark & \checkmark & \checkmark & \checkmark \\ \hline

\begin{tabular}[c]{@{}l@{}}Utilize news\\content semantic\\information.\end{tabular} & \checkmark & \checkmark &\checkmark & \checkmark &   &   & \checkmark \\ \hline
\begin{tabular}[c]{@{}l@{}}Utilize content\\conflict among\\news.\end{tabular} & \checkmark &   &   & \checkmark &   &   &   \\ \hline

\begin{tabular}[c]{@{}l@{}}Labeled training\\data not\\required.\end{tabular} & \checkmark &   &   &   & \checkmark & \checkmark &   \\ \hline
\end{tabular}

\label{table:compare_related_work}
\end{table}

\subsection{Flow Models}

According to \cite{josang2007survey}, flow models on graphs compute the score of each node by a transitive iteration on the graph or on arbitrary chains. 
The most prominent flow model is Page-Rank \cite{page1999pagerank}: it has a crucial role in the Google search engine, evaluating a web-page's score by the number of its incoming and outgoing hyperlinks. Given an initial score on the whole internet, the energy propagates iteratively through the web-page hyperlinks. As one web-page score increases, another decreases. After a full transitive iteration, the score of each page will be stable. Another popular flow model is HITS \cite{kleinberg1999hubs}. Page-Rank has only one score for each web-page, while HITS has an authority score and a hub score for each web-page, and works on a query based web-page graph which is much smaller than the whole network. HITS is good at finding authoritative pages.
A Tightly-Knit Community (TKC) is a small set of highly interconnected web pages, which can prevent HITS from finding meaningful authoritative pages. Hence Lempel et al., proposed SALSA in \cite{lempel2000stochastic} combining the random walk from Page-Rank and the bi-graph structure from HITS, thereby resolving the TKC effect. \cite{ceglowski2003semantic} proposed another flow model in the search area. It extracts a bi-graph from the data, where one side is the terms and the other the documents. For a given query, energy is distributed to the query terms, then energy flows from terms to documents, and finally documents with the highest energies are selected as the most relevant ones. 

Flow models are also popular in the trust and reputation evaluation areas. Models similar to Page-Rank include Advogato’s reputation scheme \cite{levien2004attack}, the spreading activation model \cite{ziegler2004spreading}, and
Appleseed \cite{ziegler2005propagation}. The EigenTrust model \cite{kamvar2003eigentrust} is slightly different from these three models. It works on a peer-to-peer network, and calculates a personalized trust value for each peer. For example, for a peer A, it has an initial energy, which propagates to the whole network, and the energy collected by another peer B is the trust that peer A has to peer B. Another flow model different from the above methods is SLFTD from \cite{zhang2019slftd}, where a bigraph of data providers and entities is constructed, and trust on the data provider and the discrimination ability of each entity is iteratively updated as in HITS. 

Flow models are also widely used in other areas, such as spam checking \cite{gyongyi2006link} and link fusion \cite{xi2004link}. Also the speed-up of computation is discussed in \cite{del2005fast}, and iterative propagation stop criteria is considered in \cite{berkhin2005survey}. The main idea behind these flow models is that iterative energy propagation on the graph structure and a final stable energy distribution can help us evaluate the nodes in the graph. Clearly, flow models can be adapted to many different scenarios. In our work we use it to explore fake news.

\section{Background of the Proposed Method}
\label{sec:background}

In this section we explain connections with our previous work that are relevant for our approach to fake news detection.
The two main concepts are the Inconsistency Graph (IG) based on the property of the data and the Energy Flow (EF) method. EF will be used to optimize the IG by obtaining the final energy of each node.
\subsection{Inconsistency Graph}

In our previous work \cite{xu2019incompfuse}, we introduced the concept of incompatibility probability (IP). Inconsistency  is a special case where $IP$ = 1. We also introduced the concept of an Incompatibility Graph (IG) in our recent work \cite{XU2020101508} 
to help detect noisy reports in disease data. In this paper we use IG for the 
Inconsistency Graph as we are considering only inconsistency.

The Inconsistency Graph (IG) for a set of news items contains nodes for the news items and edges for inconsistent opinions between them.
Consider what such an IG can show us:

\begin{itemize}
    \item Graphs with higher connectivity correspond to news sets with lower reliability (there is more disagreement in the opinions). 
    \item For each node, higher connectivity means higher inconsistency and lower reliability (there is more disagreement in the opinions for that node).
    \item Inconsistency with respect to less reliable nodes is better than inconsistency with more reliable nodes. 
    \item Disconnected nodes correspond to news items with the highest reliability (there is no disagreement in the opinions).
\end{itemize}
	
    \begin{figure*}
        \centering
        \begin{subfigure}[b]{1\textwidth}
            \centering
\includegraphics[width=4cm]{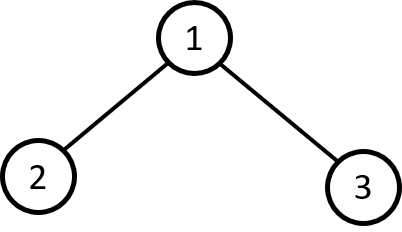}
        \end{subfigure}
        \par
        \caption{Example of an Inconsistency Graph (IG)}
	\label{fig:IG1}
    \end{figure*}	 

\begin{table}[]
\centering
\caption{The Great Moon Hoax}
\begin{tabular}{|l|l|l|}
\hline
\multicolumn{1}{|c|}{News ID} & \multicolumn{1}{c|}{News} & \multicolumn{1}{c|}{Resource} \\ \hline
1 & \begin{tabular}[c]{@{}l@{}}Herschel found evidence of life forms on\\ the moon.\end{tabular} & \begin{tabular}[c]{@{}l@{}}New York Sun, \\ Aug 1835\end{tabular} \\ \hline
2 & \begin{tabular}[c]{@{}l@{}}The news that Herschel found evidence of\\ life forms on the moon is a hoax.\end{tabular} & \begin{tabular}[c]{@{}l@{}}New York Herald, \\ Aug 1835\end{tabular} \\ \hline
3 & \begin{tabular}[c]{@{}l@{}}The news that Herschel found evidence of\\ life forms on the moon is a palpable hoax.\end{tabular} & \begin{tabular}[c]{@{}l@{}}Boston Morning Post,\\ Aug 1835\end{tabular} \\ \hline
\end{tabular}
\label{table:moon hoax}
\end{table}
	
Figure~\ref{fig:IG1} shows an example of a simple Inconsistency Graph (IG) reflecting conflicts among three news reports from the Great Moon Hoax\cite{GreatMoonhoax}. In August 1835, the New York Sun published articles saying that Sir John Herschel found evidence of life forms on the moon. Many other newspapers responded immediately with skepticism. Table ~\ref{table:moon hoax} shows three news reports, report 1 is from the New York Sun, which is in conflict with the other two reports from the New York Herald and the Boston Morning Post, respectively.

\subsection{Energy Flow method}
We can evaluate the reliability of each news item as a node in an Inconsistency Graph applying the Energy Flow method \cite{XU2020101508}. 
In this approach we inject an initial energy into each node and have it propagated to other nodes along the IG edges until the energy distribution on all nodes converges. The IG edges are also updated continuously based on the current energy distribution for each node:

\begin{center}
\begin{equation} \label{eq:ig0}
 IG(i,j)=0;   
\end{equation}
for the initial state if $Node_i$ and  $Node_j$ are consistent;
\end{center}

\begin{center}
   \begin{equation} \label{eq:ig1}
 IG(i,j)=1;     
\end{equation}
for the initial state 
        if $Node_i$ and $Node_j$ are inconsistent; 
\end{center}

\begin{center}
\begin{equation} \label{eq:ig update}
IG(i,j)=-log_{10}(\frac{Energy(j)}{\sum_{i=1}^{N} Energy(i)});
\end{equation}
during propagation if $Node_i$ and $Node_j$ are inconsistent;
\end{center}

    \begin{figure*}
        \centering
        \begin{subfigure}[b]{1\textwidth}
            \centering
\includegraphics[width=\linewidth]{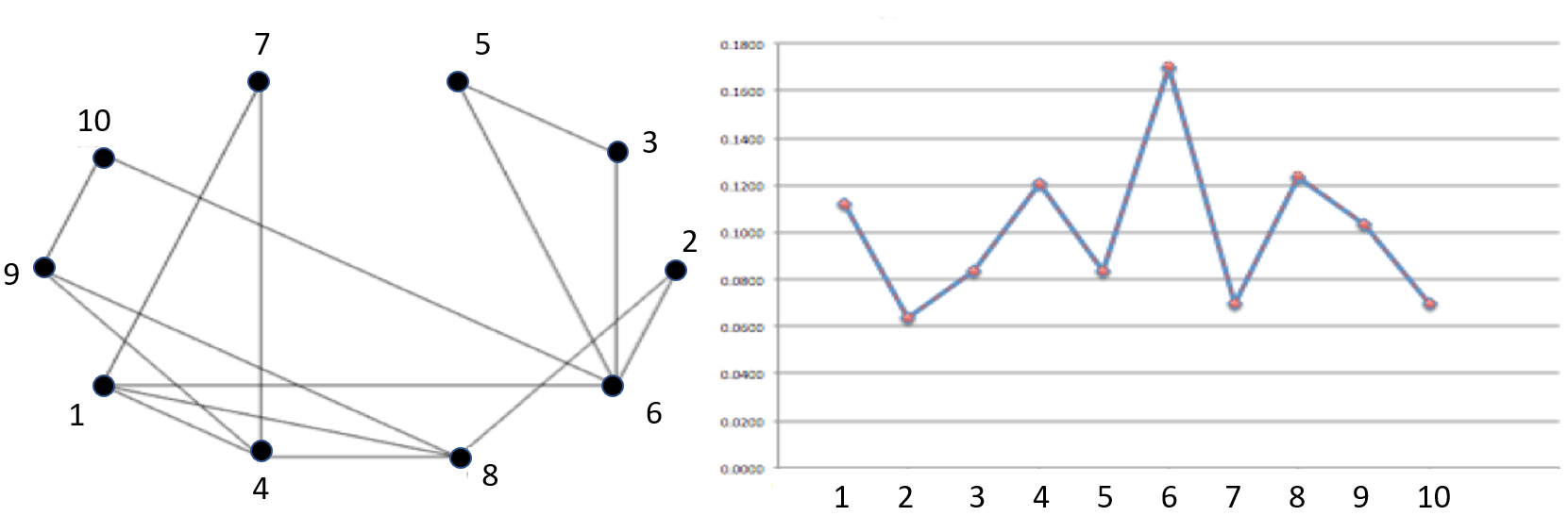}
        \end{subfigure}
        \par
\caption{Data credence assessment via energy propagation in the Inconsistency Graph}
	\label{fig:EF1}
    \end{figure*}

Nodes with  higher final energy will be considered less reliable. Figure~\ref{fig:EF1} illustrates this approach for the example of Figure~\ref{fig:IG1}. The explanation of the steps used In Figure~\ref{fig:EF1} can be found in our recent paper \cite{XU2020101508}.

After the calculation for our example, the energy of both Node 3 and Node 1 is 75, less than for Node 2, 150. Node 2 has the highest connectivity and the highest energy, so it is the least reliable node as expected. This illustrates why for each node, higher connectivity means lower reliability.

\section{FaNDS: Fake News Detection System}
\label{sec:method} 

\subsection{Inconsistency Graph (IG) creation}

The problem we study in this paper is much more complicated than the simple example in Figure~\ref{fig:IG1}. 
We use the original data from the Fake News Challenge (FNC-1) \footnote{\href{http://www.fakenewschallenge.org/}{http://www.fakenewschallenge.org/}}.
 FNC-1 focuses on the first stage of Fake News detection, called Stance Detection, to understand what other news organizations are saying about the topic. Our approach focuses on the second stage of Fake News Detection, to find which news is more likely to be fake news, based on the results of FNC-1.
 
 As our first step in the creation of the Inconsistency graph (IG), 
 we reconstructed the FNC-1 results into three tables:
 \begin{itemize}
     \item A Topic table consisting of the information of the topics,
     \item A News Article table consisting of the information of the news articles,
     \item A Stance table consisting of the relationship between a topic and news articles. The stance has four possible values: agree, disagree, discuss, and unrelated. In this paper, we consider only two of the four stances: agree and disagree. (This is how we get inconsistencies.) Therefore, a single news article either agrees or disagrees with a corresponding topic. This selection is then processed in the second step.
 \end{itemize}
 
 It should be noted that there is a many-to-many relationship between articles and topics: one news article may correspond to several topics and one topic may be related to several news articles.
 
 In the second step to generate the Inconsistency Graph (IG), we search for the inconsistent pairs of news articles. In the data set from the first step, we only have the relationship between topics and news articles but no relationship between articles. Algorithm~\ref{algo:pair} shows how to find the contradictory relationships between articles, and create the list of inconsistency pairs for the Inconsistency Graph (IG). Table~\ref{table:i-pair} shows an example of the inconsistency pairs in a Topic-News Relationship Table. In this table, after applying Algorithm~\ref{algo:pair}, the inconsistency pairs are: (1,3),(2,3),(5,6), and (2,8), using the News ID numbers.

 \IncMargin{1em}
\begin{algorithm}
\SetKwData{Left}{left}\SetKwData{This}{this}\SetKwData{Up}{up}
\SetKwFunction{Union}{Union}\SetKwFunction{FindCompress}{FindCompress}
\SetKwInOut{Input}{input}\SetKwInOut{Output}{output}

\Input{Stance Table (an example is in Table~\ref{table:t-n})}
\Output{Inconsistency pairs of articles for the Inconsistency Graph (IG)}
\BlankLine
\emph{Read the Stance Table line by line}\;
\BlankLine
\emph{Initialize a list\_inconsistent\_pair, to store the inconsistent pair of articles}\;
\BlankLine
\For{$i\leftarrow 1$ \KwTo \#\_of\_topics}{

    \emph{Initialize a list\_agree, to store the articles with agree stance}\;
    \emph{Initialize a list\_disagree, to store the articles with disagree stance}\;
    \For{$j\leftarrow 1$ \KwTo \#\_of\_articles\_related\_to\_topic\_i}{
        \If{stance\_of\_j is agree}{
            \emph{put article\_j into the list\_agree}\;
        }
        \If{stance\_of\_j is disagree}{
            \emph{put article\_j into the list\_disagree}\;
        }        
    }

    \emph{Pair each of the articles in the list\_agree with each of the articles in the list\_disagree and store them in the list\_inconsistent\_pair}\;    

}
 \caption{Creation of inconsistency pairs of articles for Inconsistency Graph (IG)}
 \label{algo:pair}
\end{algorithm}

\begin{table}
\centering
\caption{Example of Inconsistency pairs in a Topic-News Relationship Table}
\begin{tabular}{|c|c|c|}
\hline
Topic ID & News ID & Stance           \\
\hline
1    & 1    & agree                  \\
1    & 2     & agree                \\
1    & 3    & disagree                   \\
1    & 4    & discuss                      \\
2    & 5    & agree                 \\
2    & 6    & disagree                \\
3    & 2    & agree                    \\
3    & 8    & disagree                  \\
3    & 9    & discuss \\
\hline
\end{tabular}
\label{table:i-pair}
\end{table}

  After the screening of all the topics, some inconsistent article pairs may appear several times since one article may correspond to several topics. In this paper, the same article pair only counts once no matter how many times it appears.
  
 \begin{figure}
        \centering
        \begin{subfigure}[b]{1\textwidth}
            \centering
\includegraphics[width=5cm]{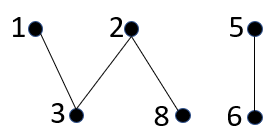}
        \end{subfigure}
        \par
\caption{Inconsistency Graph of the inconsistent article pairs}
	\label{fig:IG-small}
    \end{figure}	  
 
 After the generation of all the inconsistent article pairs, we create the Inconsistency Graph. In IG, each node represents an article. An edge between two nodes means an inconsistent opinion (the stance) between them on the same topic. Figure~\ref{fig:IG-small} shows the IG of the simple example in Table~\ref{table:i-pair}.
 
 \begin{figure*}
        \centering
        \begin{subfigure}[b]{1\textwidth}
            \centering
\includegraphics[width=\linewidth]{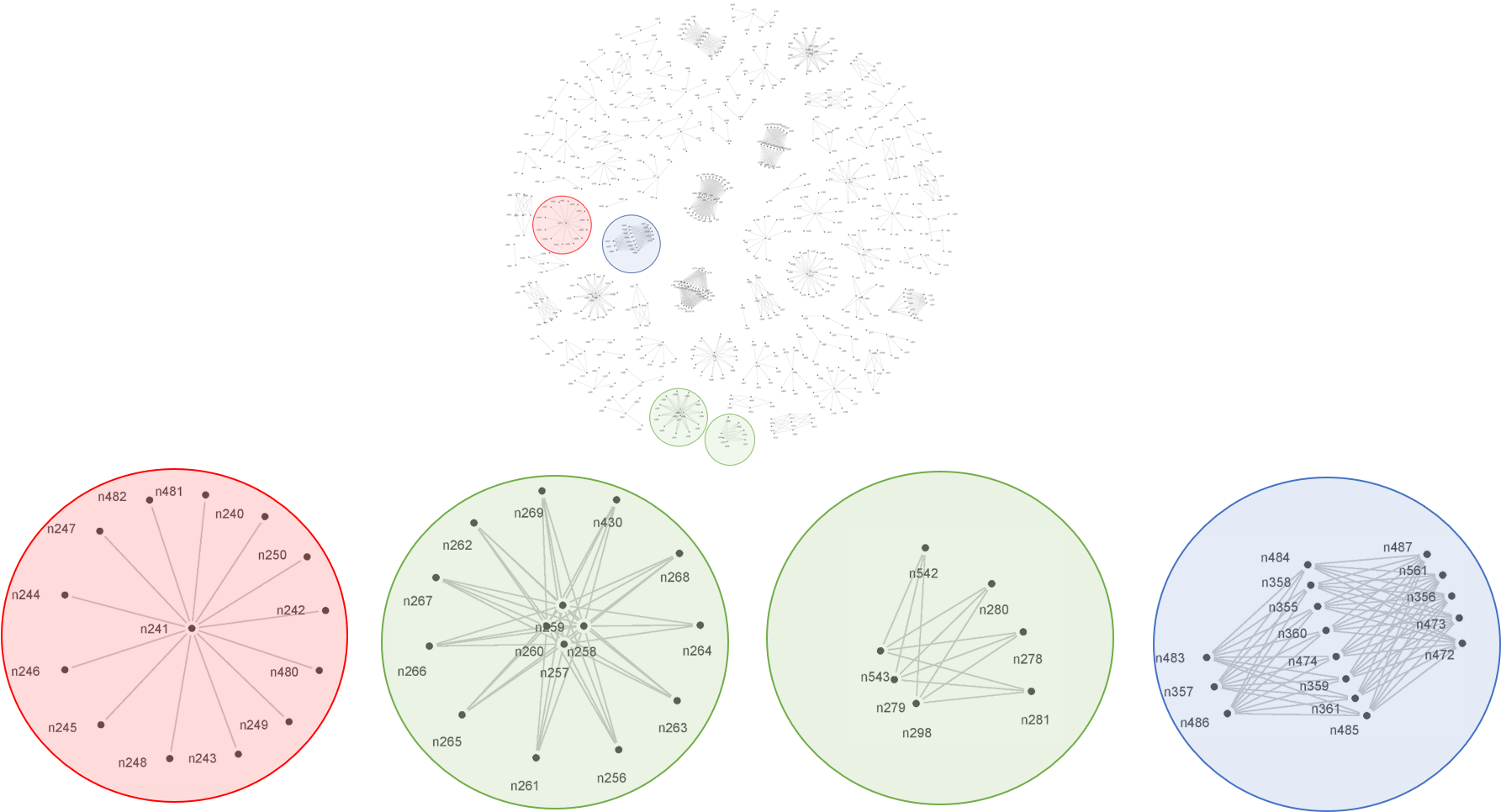}
        \end{subfigure}
        \par
\caption{Inconsistency Graph of the inconsistent article pairs}
	\label{fig:IG whole}
    \end{figure*}	  

Figure~\ref{fig:IG whole} shows the IG of a more complicated case, which consists of 586 article nodes. This Inconsistency Graph (IG) is plotted using Data-Driven Documents (D3) \footnote{\href{https://d3js.org/}{https://d3js.org/}}, which is one of the most effective frameworks to work on data visualization by utilizing a resource JavaScript library for managing documents based on data\cite{bostock2011d3}. It can be seen that the IG consists of a few isolated subgraphs with two major shapes that we call Polygon (blue on the right) and Star (red on the left). In a Polygon shape subgraph, the nodes are divided into two groups which are inconsistent with each other, and the number of nodes in each group is similar. On the other hand, in a Star shape subgraph, although the nodes are also split into two groups, the number of nodes in one group is much larger than in the other group. All other subgraphs are between these two, such as the Polygon-like subgraph (green on the right) and the Star-like subgraph (green on the left).

\subsection{Energy Flow Method via a Transformation Matrix Model}

After creating the IG, we apply the Energy Flow method to evaluate the reliability of each node. As described in detail in our previous paper \cite{XU2020101508},
the energy flow process can be represented by a Transformation Matrix Model, which is defined as:

    \begin{center}
        \begin{equation}
            \label{}
E_{n+1}=M_n*E_n              
        \end{equation}
    \end{center}
                                                          
$E$ is the energy on the nodes, and $M$ is the transformation matrix. $E_n$ is the energy after the $n^{th}$ step. After the energy converges, the equation will be

    \begin{center}
        \begin{equation}
            \label{}
E=M*E            
        \end{equation}
    \end{center}

The formation of the transformation matrix, $M$, is defined as:

    \begin{center}
        \begin{equation}
            \label{eq:M1 update}
M_{n+1}(i,j)=\left\{\begin{matrix}
1-p,\ if(i=j)\\ 
0,\ if(i\neq j \ and\ M_n(i,j)=0)\\ 
\frac{AF(j)*p}{AF^{T}*M_n}(j),\ if(i\neq j\ and\ M_n(i,j)\neq0)
\end{matrix}\right.            
        \end{equation}
    \end{center}

where

    \begin{center}
        \begin{equation}
            \label{}
AF(j)= -log_{10}(\frac{Energy(j)}{\sum_{i=1}^{N} Energy(i)})            
        \end{equation}
    \end{center}

Here $M_1$ represents the initial IG and $E_1$ the initial energy\cite{XU2020101508}.

\subsection{Relative Energy}

The magnitude of the final energy on each node depends not only on the structure of the Inconsistency Graph, but also on the magnitude of the initial energy injected into each node at the beginning. To make the measurement more uniform, we normalized the final energy to a relative energy with the range of [0, 1], as follows\cite{XU2020101508}:
    \begin{center}
        \begin{equation}
            \label{}
Energy_{relative}(j)= Energy(j)/E_{highest}      
        \end{equation}
    \end{center}
in which, $E_{highest}$ is the highest non-zero energy on the nodes. The node with the highest energy will also have the highest relative energy, 1, and will be the least reliable node with the highest inconsistency with other reports. Any report with energy 0 also has a 0 relative energy, which is a most reliable node that has no inconsistency with any other node.

\subsection{Verification of the Fake News}
\label{subsec:verify}
The nodes (news/articles) with high relative energy are considered as candidates for fake news. 

Then we have to verify that they are truly fake news. 
The challenge is that it may be difficult to know if the evidence for fake news verification from a certain source is reliable. In this paper, we applied several rules to check this:
\begin{itemize}
    \item We chose the evidence from the highly reliable media sources as rated by Media Bias/Fact Check\footnote{\href{https://www.mediabiasfactcheck.com/}{https://www.mediabiasfactcheck.com/}}, which is a widely cited source for ratings of news sources and their biases\cite{bountouridis2018explaining,baly2018predicting,fairbanks2018credibility}. In Media Bias/Fact Check, there is a Factual Reporting factor for ratings of news sources with six categories, Very High, High, Mostly Factual, Mixed, Low, and Very Low. In this paper, we only choose those with "High" or "Very High" Factual Reporting rate. 
    \item We also used Snopes\footnote{\href{https://www.snopes.com}{https://www.snopes.com}}, which is a fact-checking website and is a well-regarded reference for sorting out fake news and rumors on the internet\cite{allcott2017social,heath2016facebook,berghel2017lies,bounegru2018field}.
    \item For an article corresponding to a famous person, we also checked the related Wikipedia website. The evidence on wiki with a reliable reference is also accepted.
    \item To verify each candidate fake news item, we used multiple evidence from different reliable sources.
\end{itemize}
The details of the utilization of these rules are given in Section~\ref{sec:result ana}.

\section{Experimental Study}
\label{sec:experiments}

We start by giving the setup we used for the experiments.
That is followed by various graphs showing the effectiveness of the proposed method.

\subsection{Experimental Setup}
\label{subsec:exp setup}

We conducted our experiments on a computer with a CPU of Intel Core i7-8750H, which has 6 cores with processor clocks at between 2.2 and 4.1 GHz. The RAM of the computer is 8GBx2(16Gb) DDR4-2666(1333Hz). We implemented our method and performed the experimental study in Matlab R2018a. 

As mentioned in Section~\ref{sec:method}, the initial experimental data is from the Fake News Challenge (FNC-1) \footnote{\href{http://www.fakenewschallenge.org/}{http://www.fakenewschallenge.org/}}. FNC-1 focuses on the first stage of Fake News detection, while our approach focuses on the second stage to find which news is more likely to be fake news, based on the results of first stage.

\begin{table}
\centering
\caption{Topic-News Relationship Table Sample (75386 tuples)}
\begin{tabular}{|c|c|c|c|}
\hline
Topic ID & News ID & Stance           & Stance ID  \\
\hline
1        & 2093    & discuss          & 2          \\
1        & 186     & unrelated        & 3          \\
1        & 2367    & disagree         & 1          \\
...      & ...     & ...              & ...        \\
2        & 1704    & unrelated        & 3          \\
2        & 1994    & unrelated        & 3          \\
...      & ...     & ...              & ...        \\
2731     & 1272    & agree            & 0          \\
2731     & 2136    & unrelated        & 3          \\
\hline
\end{tabular}
\label{table:t-n}
\end{table}

In our experiments, we first followed the 3 steps in Section~\ref{sec:method} to create the Inconsistency Graph (IG). Here are the results step by step:

\begin{itemize}
    \item In the first step we found 2731 topics in the Topic Table, 2586 news items in the News Table, and 75386 tuples in the Stance Table. A sample of the records in the Stance table is shown in Table~\ref{table:t-n}.
    \item In the second step only 586 news articles survived to make up of 7244 inconsistency pairs (including repeated pairs) related to 471 topics.
    \item In the third step the Inconsistency Graph (IG), shown in Figure~\ref{fig:IG1}, was generated. 
\end{itemize}

\subsection{Relative Energy of all the nodes}

\begin{figure}[ht!]
	\centering
\includegraphics[width=8cm]{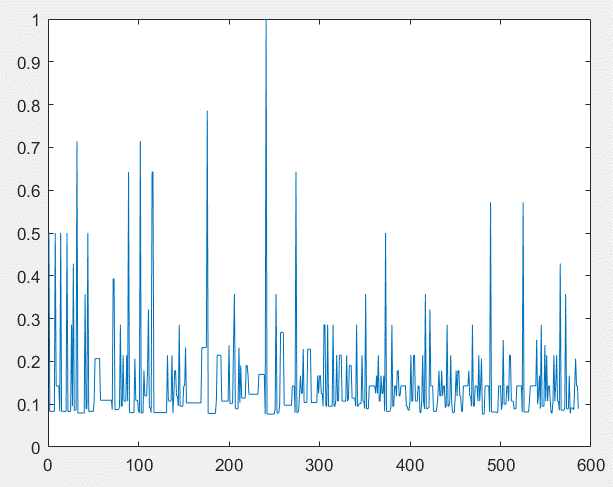}
\caption{Relative energy distribution of the 586 news items}
	\label{fig:fake news energy}
\end{figure}

After the IG was created, the Energy Flow (EF) method was applied to sort the news. Figure ~\ref{fig:fake news energy} shows the relative energy of all 586 news items. First, we can see that there is no 0 energy news, as all the nodes in the graph have at least one edge. Second, most of the nodes have relatively low energy under 0.3; only a few nodes have relatively high energy above 0.5. Since the graph shows the relative energy, the highest value is 1.

\begin{figure*}
        \centering
        \begin{subfigure}[b]{1\textwidth}
            \centering
\includegraphics[width=16cm]{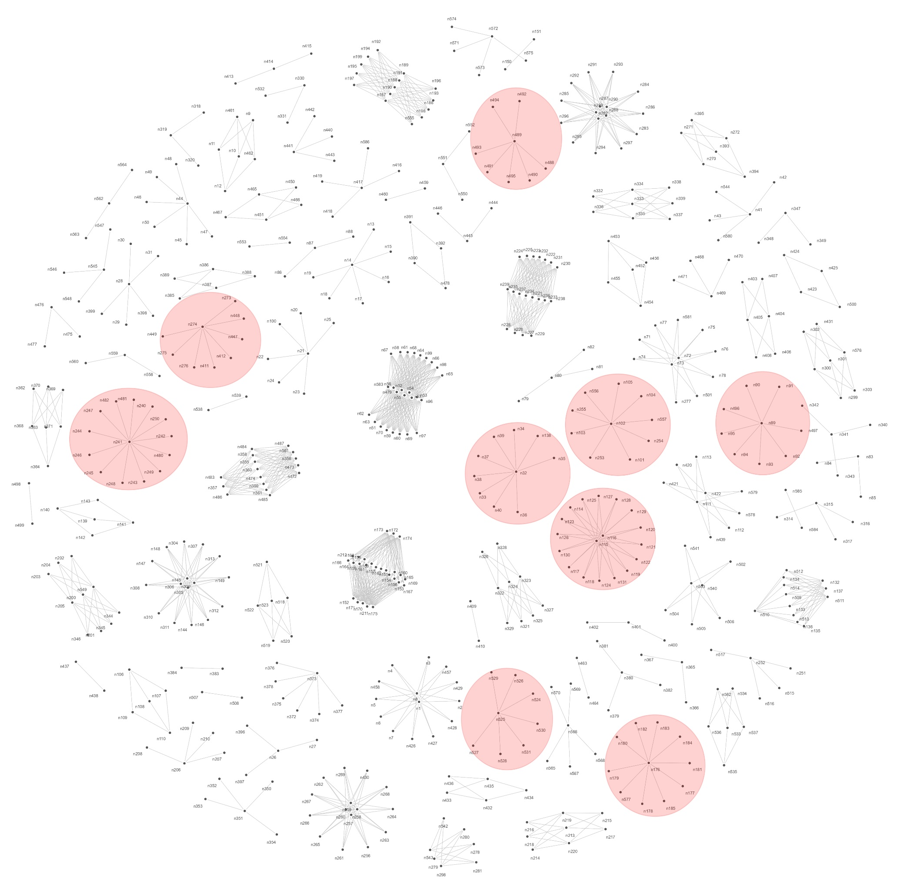}
        \end{subfigure}
        \par
\caption{Inconsistency Graph of the 586 news nodes. The red areas are the sub-graphs containing the top 10 high energy nodes }
	\label{fig:IG 586}
    \end{figure*}	 

Figure~\ref{fig:IG 586} is the Inconsistency Graph of the 586 news nodes. The red areas are the subgraphs containing the top 10 high energy nodes.

Table~\ref{table:h e news} shows some properties of the 10 nodes with the highest energy. The first column is the energy of the 10 highest energy nodes. The second column is the average energy of the nodes connected to the high energy node. These energy values are always small. The third column is the number of nodes connected to the high energy node, and the fourth column is the number of topics related to the high energy node. It can be seen that there is no direct relationship between the energy and the number of connected nodes or the number of corresponding topics, while more connected nodes with fewer corresponding topics may indicate relatively high energy.

\begin{table*}[ht!]
\centering
\caption{Properties of the 10 Nodes with the Highest Energy}
\begin{tabular}[c]{|p{10em}|p{10em}|p{7em}|p{7em}|}
\hline
Energy of high energy nodes & Energy of its connected nodes & \# of connected nodes & \# of corresponding topics \\ \hline
1 & 0.076925089                   & 13                    & 6                         \\ \hline
0.785714753                & 0.078573437                   & 10                    & 2                         \\ \hline
0.714286337                & 0.079367086                   & 9                     & 4                         \\ \hline
0.714286337                & 0.079367086                   & 9                     & 9                         \\ \hline
0.642857921                & 0.080359148                   & 8                     & 7                         \\ \hline
0.642857921                & 0.080359148                   & 16                    & 11                        \\ \hline
0.642857921                & 0.080359148                   & 16                    & 21                        \\ \hline
0.642857921                & 0.080359148                   & 8                     & 6                         \\ \hline
0.571429506                & 0.081634655                   & 7                     & 3                         \\ \hline
0.571429506                & 0.081634655                   & 7                     & 1                         \\ \hline
\end{tabular}
\label{table:h e news}
\end{table*}

\subsection{Result Analysis}
\label{sec:result ana}
In this section, we present our analysis of the top 10 highest energy articles (nodes of IG) which were detected and considered as either Fake News or not Fake News. 
The results of the analysis are listed in Table~\ref{tab:top 10 ana}:
\begin{itemize}
    \item The 1st column is the rank of the top 10 highest energy articles. 
    \item The 2nd column is the node ID in the Inconsistency Graph (IG).
    \item The 3rd column is the article ID in the Topic-News Relationship Table (see Table~\ref{table:t-n}). It is also the article ID in the FNC-1 data set.
    \item The 4rd column is the opinion of the article, either supporting a statement or opposing one.
    \item The 5th column is the type of an article, either Fake News, or not Fake News, 
    which is determined based on the evidence.
    \item The 6th column is the evidence used for the determination of the article type.
\end{itemize}
   To get reliable evidence, we carefully selected the sources of the evidence 
   as explained in Section~\ref{subsec:verify}. We gathered reliable evidence to determine the type of an article as follows:
\begin{itemize}
    \item For an article whose opinion is opposing a statement, We looked for the articles supporting this statement from  reliable media sources.
    \item For an article whose opinion is supporting a statement, we looked for the articles opposing this statement from reliable media sources.
    \item For an article whose opinion is supporting a statement, we searched for any article with the same opinion and labeled as fake in Snopes\footnote{\href{https://www.snopes.com}{https://www.snopes.com}}.
    \item For the selection of the reliable media sources, we only chose those with "High" or "Very High" factual reporting rate by Media Bias/Fact Check\footnote{\href{https://www.mediabiasfactcheck.com}{https://www.mediabiasfactcheck.com}}.
    \item For an article corresponding to a famous person, we also checked the related Wikipedia website. The evidence on wiki with a reliable reference was also accepted.
    \item For each article, at least two pieces of evidence from different reliable sources were used and listed. 
    
\end{itemize}
Based on our analysis, all top 10 articles from the total of 586 news/articles were fake news. Although our method can also be used to detect rumor corrections, that is not the case for any of the top 10 articles. We also found, as expected, that the original source of the fake news usually also has a bad reputation for trustworthiness. For example, the fake news about Graffiti Artist Banksy (TOP 1) was originally from a media source named National Report\footnote{\href{http://nationalreport.net/banksy-arrested-identity-revealed/}{http://nationalreport.net/banksy-arrested-identity-revealed/}}, which is rated as SATIRE by Media Bias/Fact Check. A similar observation can also be found in ~\cite{bounegru2018field}. 

\begin{table}[]
\small
    \centering
    \caption{Analysis of the top 10 of highest energy nodes (news items)} 
    \begin{threeparttable}
    \begin{tabular}{|c|c|c|c|c|c|}
    \hline
        \begin{tabular}{@{}c@{}}
            Rank
        \end{tabular}
         & 
        \begin{tabular}{@{}c@{}}
            Node\\ID
        \end{tabular}         
         &
        \begin{tabular}{@{}c@{}}
            Article\\ID
        \end{tabular}           
         &
        \begin{tabular}{@{}c@{}}
            News Item
        \end{tabular}           
         &
        \begin{tabular}{@{}c@{}}
            Type
        \end{tabular}           
         &
        \begin{tabular}{@{}c@{}}
            Evidence*
        \end{tabular}        
    \\\hline
    1 & 241 & 974 &
         \begin{tabular}{@{}p{18em}@{}}
            Graffiti Artist Banksy was arrested in London in 2014
        \end{tabular} &
        \begin{tabular}{@{}c@{}}
            Fake\\News              
        \end{tabular} &
         \begin{tabular}{@{}c@{}}
            Banksy's Wiki$^{1.1}$,\\Snopes$^{1.2}$ 
        \end{tabular}
   \\\hline
   2 & 176 & 313 &
         \begin{tabular}{@{}p{18em}@{}}
            Ahmed Godane, the head of the Somali Islamist
            militant group al Shabaab, was NOT killed in 
            a U.S. strike
        \end{tabular}&
        \begin{tabular}{@{}c@{}}
            Fake\\News              
        \end{tabular} &
         \begin{tabular}{@{}c@{}}
            Godane's Wiki$^{2.1}$,\\Reuters$^{2.2}$
        \end{tabular}
    \\\hline          

   3 & 32 & 736 &
         \begin{tabular}{@{}p{18em}@{}}
            Pregnant Woman Losing Eye After St. Louis Cops Shoot Bean-Bag Round is NOT true
        \end{tabular}&
        \begin{tabular}{@{}c@{}}
            Fake\\News              
        \end{tabular} &
         \begin{tabular}{@{}c@{}}
            NBC News$^{3.1}$\\The Guardian$^{3.2}$ 
        \end{tabular}
    \\\hline

   4 & 102 & 2125 &
         \begin{tabular}{@{}p{18em}@{}}
            Seven teenage girls getting pregnant on school trip is NOT true
        \end{tabular}&
        \begin{tabular}{@{}c@{}}
            Fake\\News              
        \end{tabular} &
         \begin{tabular}{@{}c@{}}
            Newser$^{4.1}$,\\News.com.au$^{4.2}$
        \end{tabular}
    \\\hline   

   5 & 89 & 1372 &
         \begin{tabular}{@{}p{18em}@{}}
            11 aircraft are missing from terrorist-held Tripoli airport ahead of 9/11 anniversary
        \end{tabular}&
        \begin{tabular}{@{}c@{}}
            Fake\\News              
        \end{tabular} &
         \begin{tabular}{@{}c@{}}
            Snopes$^{5.1}$, \\USA Today$^{5.2}$
        \end{tabular}
    \\\hline   

   6** & 115 & 1451 &
         \begin{tabular}{@{}p{18em}@{}}
            A Justin Bieber Ringtone saving A Man's Life Who Was Being Attacked By A Bear is NOT true
        \end{tabular}&
        \begin{tabular}{@{}c@{}}
            Fake\\News              
        \end{tabular} &
         \begin{tabular}{@{}c@{}}
            National Post$^{6.1}$,\\The Independent$^{6.2}$ 
        \end{tabular}
    \\\hline   
    
   7** & 116 & 2373 &
         \begin{tabular}{@{}p{18em}@{}}
            A Justin Bieber Ringtone Just Saving A Man's Life Who Was Being Attacked By A Bear is NOT true
        \end{tabular}&
        \begin{tabular}{@{}c@{}}
            Fake\\News              
        \end{tabular} &
         \begin{tabular}{@{}c@{}}
            National Post$^{7.1}$,\\The Independent$^{7.2}$ 
        \end{tabular}
    \\\hline   
    
    8 & 274 & 1862 &
         \begin{tabular}{@{}p{18em}@{}}
             Gill Rosenberg former IDF soldier was captured by ISIS
        \end{tabular}&
        \begin{tabular}{@{}c@{}}
            Fake\\News              
        \end{tabular} &
         \begin{tabular}{@{}c@{}}
            Haaretz$^{8.1}$,\\The Jerusalem Post$^{8.2}$,\\Gill Rosenberg' wiki$^{8.3}$
        \end{tabular}
    \\\hline      
    
    9 & 489 & 633 &
         \begin{tabular}{@{}p{18em}@{}}
            Rumor debunked: RoboCop-style robots are NOT patrolling Microsoft's campus
        \end{tabular}&
        \begin{tabular}{@{}c@{}}
            Fake\\News              
        \end{tabular} &
         \begin{tabular}{@{}c@{}}
            MIT Technology \\Review$^{9.1}$,\\Business Insider$^{9.2}$
        \end{tabular}
    \\\hline      

    10 & 525 & 383 &
         \begin{tabular}{@{}p{18em}@{}}
            No, it's not Tiger Woods selling an island in Lake Mälaren
        \end{tabular}&
        \begin{tabular}{@{}c@{}}
            Fake\\News              
        \end{tabular} &
         \begin{tabular}{@{}c@{}}
            Business Insider$^{10.1}$,\\Stuff$^{10.2}$ 
        \end{tabular}
    \\\hline

    \end{tabular}

    \begin{tablenotes}
      \tiny
     \item *: Evidence selected from wiki with evidence references, Snopes(\href{https://www.snopes.com}{https://www.snopes.com}), or the high reputation sources based on the rates from Media Bias/Fact Check (\href{https://www.mediabiasfactcheck.com/}{https://www.mediabiasfactcheck.com/})
     \item **: 6 and 7 are two articles with the same opinion.
     \item 1.1: \href{https://en.m.wikipedia.org/wiki/Banksy}{https://en.m.wikipedia.org/wiki/Banksy}
     \item 1.2: \href{https://www.snopes.com/fact-check/graffiti-artist-banksy-arrested-london-identity-revealed/}{https://www.snopes.com/fact-check/graffiti-artist-banksy-arrested-london-identity-\\revealed/}
     \item 2.1: \href{https://en.wikipedia.org/wiki/Ahmed_Abdi_Godane#cite_note-Ubaidah-5}{https://en.wikipedia.org/wiki/Ahmed\_Abdi\_Godane\#cite\_note-Ubaidah-5} 
     \item 2.2: \href{https://www.reuters.com/article/us-somalia-usa-islamist/u-s-confirms-death-of-al-shabaab-leader-godane-in-somalia-strike-idUSKBN0H01OO20140905}{https://www.reuters.com/article/us-somalia-usa-islamist/u-s-confirms-death-of-al-\\shabaab-leader-godane-in-somalia-strike-idUSKBN0H01OO20140905} 
     \item 3.1: \href{https://www.nbcnews.com/news/us-news/pregnant-woman-loses-eye-after-st-louis-cops-shoot-bean-n257876}{https://www.nbcnews.com/news/us-news/pregnant-woman-loses-eye-after-st-louis-cops-\\shoot-bean-n257876}
     \item 3.2: \href{https://www.theguardian.com/us-news/2014/nov/29/ferguson-police-woman-blinded-one-eye}{https://www.theguardian.com/us-news/2014/nov/29/ferguson-police-woman-blinded-\\one-eye}
     \item 4.1: \href{https://www.newser.com/story/200321/7-teens-come-home-pregnant-from-school-trip.html}{https://www.newser.com/story/200321/7-teens-come-home-pregnant-from-school-\\trip.html}
     \item 4.2: \href{https://www.news.com.au/lifestyle/real-life/wtf/seven-girls-fall-pregnant-after-five-day-school\\-trip-in-bosnia-and-herzegovina/news-story/9572be83cf02f5a39783416f4ec6ac33}{https://www.news.com.au/lifestyle/real-life/wtf/seven-girls-fall-pregnant-after-\\five-day-school\\-trip-in-bosnia-and-herzegovina/news-story/9572be83cf02f5a39783416f4ec6ac33}
     \item 5.1: \href{https://www.snopes.com/fact-check/missing-planes/}{https://www.snopes.com/fact-check/missing-planes/}
     \item 5.2: \href{https://www.usatoday.com/story/news/world/2014/09/04/libya-missing-planes-sept-11/15059169/}{https://www.usatoday.com/story/news/world/2014/09/04/libya-missing-planes-sept-\\11/15059169/}
     \item 6.1: \href{https://nationalpost.com/scene/justin-bieber-song-baby-saves-russian-fisherman-from-bear-attack}{https://nationalpost.com/scene/justin-bieber-song-baby-saves-russian-fisherman-\\from-bear-attack}
     \item 6.2: \href{https://www.independent.co.uk/news/people/justin-bieber-saved-a-man-from-being-\\mauled-by-a-bear-sort-of-9651054.html}{https://www.independent.co.uk/news/people/justin-bieber-saved-a-man-from-being-\\mauled-by-a-bear-sort-of-9651054.html}
     \item 7.1: \href{https://nationalpost.com/scene/justin-bieber-song-baby-saves-russian-fisherman-from-bear-attack}{https://nationalpost.com/scene/justin-bieber-song-baby-saves-russian-fisherman-\\from-bear-attack}     
     \item 7.2: \href{https://www.independent.co.uk/news/people/justin-bieber-saved-a-man-from-being-\\mauled-by-a-bear-sort-of-9651054.html}{https://www.independent.co.uk/news/people/justin-bieber-saved-a-man-from-being-\\mauled-by-a-bear-sort-of-9651054.html} 
     \item 8.1: \href{https://www.haaretz.com/israel-checking-reports-isis-captured-israeli-1.5338025}{https://www.haaretz.com/israel-checking-reports-isis-captured-israeli-1.5338025}
     \item 8.2: \href{https://www.jpost.com/Middle-East/Israeli-Canadian-woman-reportedly-kidnapped-by-ISIS-makes-contact-Im-totally-safe-and-secure-383371}{https://www.jpost.com/Middle-East/Israeli-Canadian-woman-reportedly-kidnapped-by-ISIS-makes-contact-Im-totally-safe-and-secure-383371}
     \item 8.3: \href{https://en.wikipedia.org/wiki/Gill_Rosenberg}{https://en.wikipedia.org/wiki/Gill\_Rosenberg}
     \item 9.1: \href{https://www.technologyreview.com/s/532431/rise-of-the-robot-security-guards/}{https://www.technologyreview.com/s/532431/rise-of-the-robot-security-guards/}
     \item 9.2: \href{https://www.businessinsider.com/knightscope-security-robots-microsoft-uber-2017-5}{https://www.businessinsider.com/knightscope-security-robots-microsoft-uber-2017-5}
     \item 10.1: \href{https://www.businessinsider.com/tiger-woods-former-private-island-on-sale-2015-1}{https://www.businessinsider.com/tiger-woods-former-private-island-on-sale-2015-1}
     \item 10.2: \href{https://www.stuff.co.nz/life-style/home-property/64789215/tiger-woods-selling-private-island}{https://www.stuff.co.nz/life-style/home-property/64789215/tiger-woods-selling-private-island}
    \end{tablenotes}    
\end{threeparttable}
\label{tab:top 10 ana}
\end{table}

\subsection{Comparison and Sensitivity Analysis with related methods}

 In the previous section, we represented the performance of our FaNDS method for detecting fake news and the correction of rumors for the FNC-1 data. In this section, in order to do a direct comparison of the performance of our method with others, we applied the following methods: Max voting by count, Max voting by percentage, Hub Authority, and FaNDS to detect the fake news items.

 \begin{table}[]
 \centering
 \begin{threeparttable}
\caption{Top 10 fake news articles by different methods}
\label{tab:compare}
\begin{tabular}{|c|c|c|c|c|}
\hline
\begin{tabular}{@{}c@{}}
Rank \\Top 10
\end{tabular}

& 
\begin{tabular}{@{}c@{}}
Max Voting\\ by Count*
\end{tabular}
 &
\begin{tabular}{@{}c@{}}
 Max Voting by\\ Percentage*
\end{tabular}
&
\begin{tabular}{@{}c@{}}
Hub Authority*
\end{tabular}
& 
\begin{tabular}{@{}c@{}}
FaNDS* 
\end{tabular} \\\hline
1 & \cellcolor[HTML]{C7C5C5} 152 &\cellcolor[HTML]{C7C5C5} 241 &\cellcolor[HTML]{C7C5C5} 150 &\cellcolor[HTML]{C7C5C5} 241 \\
2 & \cellcolor[HTML]{C7C5C5} 170 &\cellcolor[HTML]{E9E9E9} 176 &\cellcolor[HTML]{C7C5C5} 151 &\cellcolor[HTML]{E9E9E9} 176 \\
3 & \cellcolor[HTML]{C7C5C5} 171 &\cellcolor[HTML]{C7C5C5} 32 &\cellcolor[HTML]{C7C5C5} 383 &\cellcolor[HTML]{C7C5C5} 32 \\
4 & \cellcolor[HTML]{C7C5C5} 172 &\cellcolor[HTML]{C7C5C5} 102 &\cellcolor[HTML]{C7C5C5} 384 &\cellcolor[HTML]{C7C5C5} 102 \\
5 & \cellcolor[HTML]{C7C5C5} 173 &\cellcolor[HTML]{E9E9E9} 89 &\cellcolor[HTML]{C7C5C5} 409 &\cellcolor[HTML]{E9E9E9} 89 \\
6 & \cellcolor[HTML]{C7C5C5} 174 &\cellcolor[HTML]{E9E9E9} 115 &\cellcolor[HTML]{C7C5C5} 410 &\cellcolor[HTML]{E9E9E9} 115 \\
7 & \cellcolor[HTML]{C7C5C5} 175 &\cellcolor[HTML]{E9E9E9} 116 &\cellcolor[HTML]{C7C5C5} 437 &\cellcolor[HTML]{E9E9E9} 116 \\
8 & \cellcolor[HTML]{C7C5C5} 211 &\cellcolor[HTML]{E9E9E9} 274 &\cellcolor[HTML]{C7C5C5} 438 &\cellcolor[HTML]{E9E9E9} 274 \\
9 & \cellcolor[HTML]{E9E9E9} 52 &\cellcolor[HTML]{C7C5C5} 489 &\cellcolor[HTML]{C7C5C5} 459 &\cellcolor[HTML]{C7C5C5} 489 \\
10 & \cellcolor[HTML]{E9E9E9} 53 &\cellcolor[HTML]{C7C5C5} 525 &\cellcolor[HTML]{C7C5C5} 460 &\cellcolor[HTML]{C7C5C5} 525 \\\hline
\end{tabular}
    \begin{tablenotes}
      \small
     \item * The group of cells with the same color has the same score. The Hub Authority method has the worst sensitivity in the sense that it gave the top 22 items the same score.
    \end{tablenotes}
  \end{threeparttable}
\end{table}
 
 Table~\ref{tab:compare} shows the results:
 \begin{itemize}
     \item  FaNDS had the same top 10 fake news articles as the Max Voting method by Percentage. 
     \item Max Voting by Count gave a different list of the top 10 fake news articles.
     \item The Hub Authority method is not sensitive to the fake news detection task as there are 22 articles with the same highest score (only 10 of which are listed in the table).
 \end{itemize}
 
     \begin{figure*}
        \centering
        \begin{subfigure}[b]{1\textwidth}
            \centering
\includegraphics[width=\linewidth] {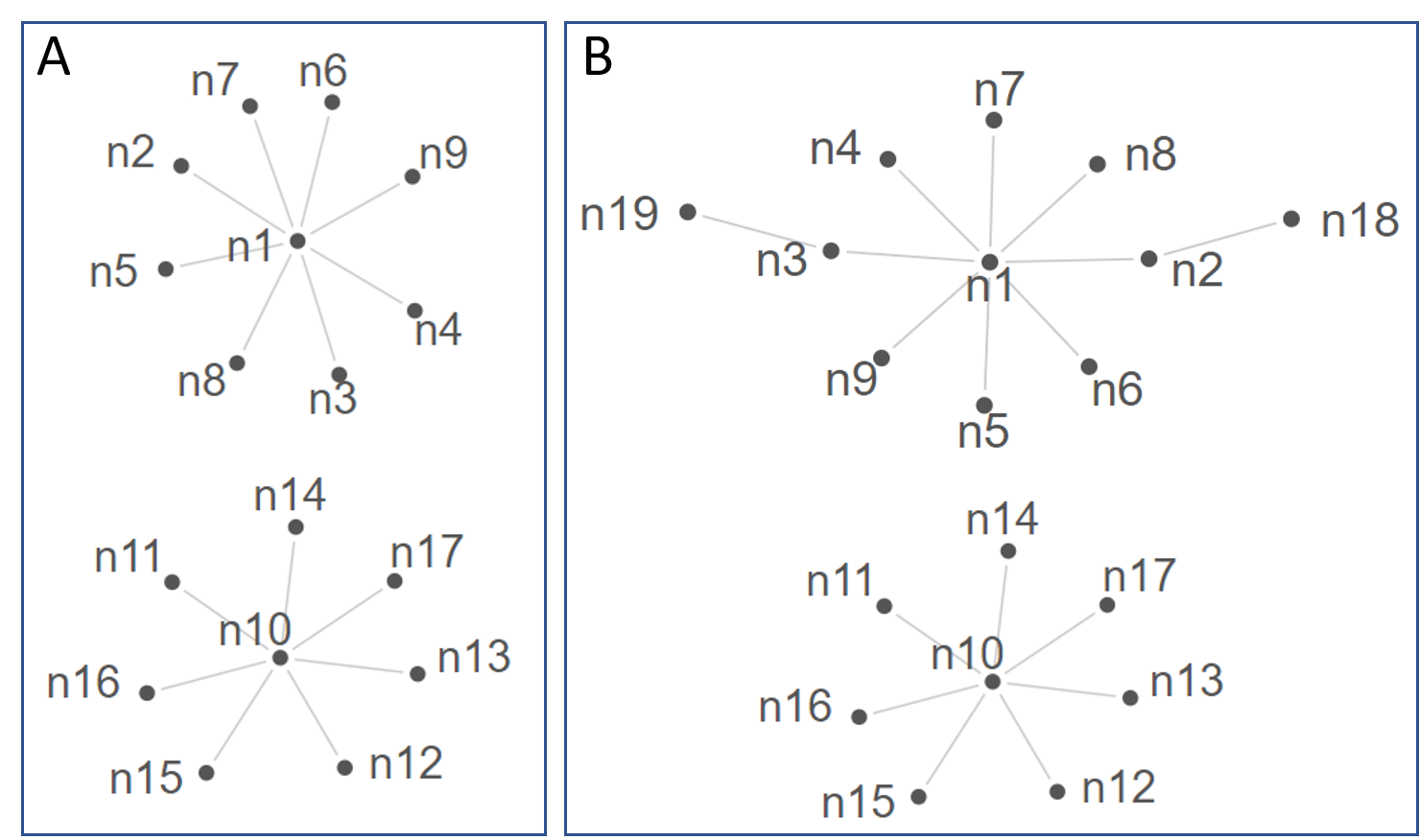}
        \end{subfigure}
        \par
\caption{Inconsistency Graph of 17 nodes (A) and of 18 nodes (B)}
	\label{fig:graph 2 graph}
    \end{figure*}	 
    
  We then considered a little bit more complicated situation where one article may include opinions related to more than one topic. 
  
  Figure~\ref{fig:graph 2 graph} shows an example. In this figure, 
  \begin{itemize}
      \item Graph A on the left with two isolated subgraphs can be considered as a portion of Figure~\ref{fig:IG 586}. One subgraph (top) has 9 nodes, in which n1 is inconsistent with all the other 8 nodes. The node n1 is same as the node n274 in Figure~\ref{fig:IG 586}, which is eighth in Table~\ref{tab:compare}. The other subgraph (bottom) has 8 nodes, in which n10 is inconsistent with all the other 7 nodes. The node n10 is same as n489 in Figure~\ref{fig:IG 586}, which is the ninth in  Table~\ref{tab:compare}.
      \item Graph B in Figure~\ref{fig:graph 2 graph} has 2 more nodes, n18 and n19, than Graph A. The node n18 forms a new inconsistency edge with n2, while the node n19 forms a new inconsistency edge with n3. The new connections represent a situation where an article (n3 or n2) may consist of more than one opinion and is related to more than one topic. It is clear that the nodes n1 to n9 are related to the same topic; n3 and n19 are related to another topic; and n2 and n18 are related to a third topic.
  \end{itemize}
  
    Table~\ref{tab:compare 2 1} shows the rank of all 17 nodes in Graph A, and Table~\ref{tab:compare 2 2} shows the rank of all 19 nodes in Graph B. It can be seen that:
    \begin{itemize}
        \item Overall, FaNDS is more sensitive in that it provides more groups of nodes than the other methods, especially in Graph B, the more complicated case. And the results from FaNDS are more reasonable than the other methods based on the assumptions. 
        \item In Graph A, the order of the top 2 nodes from Max Voting and FaNDS are the same: n1 is on top, followed by n10. The Hub Authority method gives the opposite order, but that is not reasonable, since Node 1 is more inconsistent than Node 10.
        \item However, in Graph B, the order of the top 2 nodes should be different. FaNDS gives the expected order, as introducing nodes 18 and 19 into the graph reduces the final energy on n1 to make it drop from first to second. The reason is that the existence of n18 and n19 reduces the reliability of n2 and n3 directly, and so it increases the reliability of n1 indirectly. On the other hand, the Max Voting methods are not sensitive to this kind of change.
        \item FaNDS also performs better on the rank of the nodes other than Node 1 and Node 10, especially in Graph B.
    \end{itemize}

\begin{table}[]
\centering
\caption{Rank of nodes in Graph A}
\label{tab:compare 2 1}
\begin{tabular}{|c|c|c|c|c|}
\hline
Rank
&
\begin{tabular}{@{}c@{}}
Max Voting\\ by Count
\end{tabular}
 &
\begin{tabular}{@{}c@{}}
 Max Voting by\\ Percentage
\end{tabular}
&
\begin{tabular}{@{}c@{}}
Hub Authority
\end{tabular}
& 
\begin{tabular}{@{}c@{}}
FaNDS 
\end{tabular}
\\\hline
1 &\cellcolor[HTML]{C7C5C5} 1 &\cellcolor[HTML]{C7C5C5} 1 &\cellcolor[HTML]{C7C5C5} 10 &\cellcolor[HTML]{C7C5C5} 1 \\
2 &\cellcolor[HTML]{E9E9E9} 10 &\cellcolor[HTML]{E9E9E9} 10 &\cellcolor[HTML]{E9E9E9} 1 &\cellcolor[HTML]{E9E9E9} 10 \\
3 &\cellcolor[HTML]{C7C5C5} 2 &\cellcolor[HTML]{C7C5C5} 11 &\cellcolor[HTML]{C7C5C5} 2 &\cellcolor[HTML]{C7C5C5} 11 \\
4 &\cellcolor[HTML]{C7C5C5} 3 &\cellcolor[HTML]{C7C5C5} 12 &\cellcolor[HTML]{C7C5C5} 3 &\cellcolor[HTML]{C7C5C5} 12 \\
5 &\cellcolor[HTML]{C7C5C5} 4 &\cellcolor[HTML]{C7C5C5} 13 &\cellcolor[HTML]{C7C5C5} 4 &\cellcolor[HTML]{C7C5C5} 13 \\
6 &\cellcolor[HTML]{C7C5C5} 5 &\cellcolor[HTML]{C7C5C5} 14 &\cellcolor[HTML]{C7C5C5} 5 &\cellcolor[HTML]{C7C5C5} 14 \\
7 &\cellcolor[HTML]{C7C5C5} 6 &\cellcolor[HTML]{C7C5C5} 15 &\cellcolor[HTML]{C7C5C5} 6 &\cellcolor[HTML]{C7C5C5} 15 \\
8 &\cellcolor[HTML]{C7C5C5} 7 &\cellcolor[HTML]{C7C5C5} 16 &\cellcolor[HTML]{C7C5C5} 7 &\cellcolor[HTML]{C7C5C5} 16 \\
9 &\cellcolor[HTML]{C7C5C5} 8 &\cellcolor[HTML]{C7C5C5} 17 &\cellcolor[HTML]{C7C5C5} 8 &\cellcolor[HTML]{C7C5C5} 17 \\
10 &\cellcolor[HTML]{C7C5C5} 9 &\cellcolor[HTML]{E9E9E9} 9 &\cellcolor[HTML]{C7C5C5} 9 &\cellcolor[HTML]{E9E9E9} 2 \\
11 &\cellcolor[HTML]{C7C5C5} 11 &\cellcolor[HTML]{E9E9E9} 8 &\cellcolor[HTML]{E9E9E9} 11 &\cellcolor[HTML]{E9E9E9} 3 \\
12 &\cellcolor[HTML]{C7C5C5} 12 &\cellcolor[HTML]{E9E9E9} 2 &\cellcolor[HTML]{E9E9E9} 12 &\cellcolor[HTML]{E9E9E9} 4 \\
13 &\cellcolor[HTML]{C7C5C5} 13 &\cellcolor[HTML]{E9E9E9} 3 &\cellcolor[HTML]{E9E9E9} 13 &\cellcolor[HTML]{E9E9E9} 5 \\
14 &\cellcolor[HTML]{C7C5C5} 14 &\cellcolor[HTML]{E9E9E9} 4 &\cellcolor[HTML]{E9E9E9} 14 &\cellcolor[HTML]{E9E9E9} 6 \\
15 &\cellcolor[HTML]{C7C5C5} 15 &\cellcolor[HTML]{E9E9E9} 5 &\cellcolor[HTML]{E9E9E9} 15 &\cellcolor[HTML]{E9E9E9} 7 \\
16 &\cellcolor[HTML]{C7C5C5} 16 &\cellcolor[HTML]{E9E9E9} 6 &\cellcolor[HTML]{E9E9E9} 16 &\cellcolor[HTML]{E9E9E9} 8 \\
17 &\cellcolor[HTML]{C7C5C5} 17 &\cellcolor[HTML]{E9E9E9} 7 &\cellcolor[HTML]{E9E9E9} 17 &\cellcolor[HTML]{E9E9E9} 9 \\\hline
\end{tabular}
\end{table}

\begin{table}
\centering
\caption{Rank of nodes in Graph B}
\label{tab:compare 2 2}
\begin{tabular}{|c|c|c|c|c|}
\hline
Rank & 
\begin{tabular}{@{}c@{}}
Max Voting\\ by Count
\end{tabular}
 &
\begin{tabular}{@{}c@{}}
 Max Voting by\\ Percentage
\end{tabular}
&
\begin{tabular}{@{}c@{}}
Hub Authority
\end{tabular}
& 
\begin{tabular}{@{}c@{}}
FaNDS 
\end{tabular}
\\\hline
1 &\cellcolor[HTML]{C7C5C5} 1 &\cellcolor[HTML]{C7C5C5} 1 &\cellcolor[HTML]{C7C5C5} 10 &\cellcolor[HTML]{C7C5C5} 10 \\
2 &\cellcolor[HTML]{E9E9E9} 10 &\cellcolor[HTML]{E9E9E9} 10 &\cellcolor[HTML]{E9E9E9} 1 &\cellcolor[HTML]{E9E9E9} 1 \\
3 &\cellcolor[HTML]{C7C5C5} 2 &\cellcolor[HTML]{C7C5C5} 2 &\cellcolor[HTML]{C7C5C5} 11 &\cellcolor[HTML]{C7C5C5} 2 \\
4 &\cellcolor[HTML]{C7C5C5} 3 &\cellcolor[HTML]{C7C5C5} 3 &\cellcolor[HTML]{C7C5C5} 12 &\cellcolor[HTML]{C7C5C5} 3 \\
5 &\cellcolor[HTML]{E9E9E9} 4 &\cellcolor[HTML]{E9E9E9} 11 &\cellcolor[HTML]{C7C5C5} 13 &\cellcolor[HTML]{E9E9E9} 18 \\
6 &\cellcolor[HTML]{E9E9E9} 5 &\cellcolor[HTML]{E9E9E9} 12 &\cellcolor[HTML]{C7C5C5} 14 &\cellcolor[HTML]{E9E9E9} 19 \\
7 &\cellcolor[HTML]{E9E9E9} 6 &\cellcolor[HTML]{E9E9E9} 13 &\cellcolor[HTML]{C7C5C5} 15 &\cellcolor[HTML]{C7C5C5} 11 \\
8 &\cellcolor[HTML]{E9E9E9} 7 &\cellcolor[HTML]{E9E9E9} 14 &\cellcolor[HTML]{C7C5C5} 16 &\cellcolor[HTML]{C7C5C5} 12 \\
9 &\cellcolor[HTML]{E9E9E9} 8 &\cellcolor[HTML]{E9E9E9} 15 &\cellcolor[HTML]{C7C5C5} 17 &\cellcolor[HTML]{C7C5C5} 13 \\
10 &\cellcolor[HTML]{E9E9E9} 9 &\cellcolor[HTML]{E9E9E9} 16 &\cellcolor[HTML]{E9E9E9} 2 &\cellcolor[HTML]{C7C5C5} 14 \\
11 &\cellcolor[HTML]{E9E9E9} 11 &\cellcolor[HTML]{E9E9E9} 17 &\cellcolor[HTML]{E9E9E9} 3 &\cellcolor[HTML]{C7C5C5} 15 \\
12 &\cellcolor[HTML]{E9E9E9} 12 &\cellcolor[HTML]{C7C5C5} 18 &\cellcolor[HTML]{C7C5C5} 18 &\cellcolor[HTML]{C7C5C5} 16 \\
13 &\cellcolor[HTML]{E9E9E9} 13 &\cellcolor[HTML]{C7C5C5} 19 &\cellcolor[HTML]{C7C5C5} 19 &\cellcolor[HTML]{C7C5C5} 17 \\
14 &\cellcolor[HTML]{E9E9E9} 14 &\cellcolor[HTML]{C7C5C5} 4 &\cellcolor[HTML]{C7C5C5} 4 &\cellcolor[HTML]{E9E9E9} 4 \\
15 &\cellcolor[HTML]{E9E9E9} 15 &\cellcolor[HTML]{C7C5C5} 5 &\cellcolor[HTML]{C7C5C5} 5 &\cellcolor[HTML]{E9E9E9} 5 \\
16 &\cellcolor[HTML]{E9E9E9} 16 &\cellcolor[HTML]{C7C5C5} 6 &\cellcolor[HTML]{C7C5C5} 6 &\cellcolor[HTML]{E9E9E9} 6 \\
17 &\cellcolor[HTML]{E9E9E9} 17 &\cellcolor[HTML]{C7C5C5} 7 &\cellcolor[HTML]{C7C5C5} 7 &\cellcolor[HTML]{E9E9E9} 7 \\
18 &\cellcolor[HTML]{E9E9E9} 18 &\cellcolor[HTML]{C7C5C5} 8 &\cellcolor[HTML]{C7C5C5} 8 &\cellcolor[HTML]{E9E9E9} 8 \\
19 &\cellcolor[HTML]{E9E9E9} 19 &\cellcolor[HTML]{C7C5C5} 9 &\cellcolor[HTML]{C7C5C5} 9 &\cellcolor[HTML]{E9E9E9} 9 \\\hline
\end{tabular}
\end{table}

 \begin{table}[]
 \centering
  \begin{threeparttable}
\caption{Different performance of different methods on a 10-node complicated graph }
\label{tab:compare3}
\begin{tabular}{|l|l|l|l|l|}
\hline
Rank & 
\begin{tabular}{@{}c@{}}
Max Voting\\ by Count
\end{tabular}
 &
\begin{tabular}{@{}c@{}}
 Max Voting by\\ Percentage
\end{tabular}
&
\begin{tabular}{@{}c@{}}
Hub Authority
\end{tabular}
& 
\begin{tabular}{@{}c@{}}
FaNDS* 
\end{tabular}
\\
\hline
1 &\cellcolor[HTML]{C7C5C5} 6 &\cellcolor[HTML]{C7C5C5} 6 &\cellcolor[HTML]{C7C5C5} 1 &\cellcolor[HTML]{C7C5C5} 6 \\
2 &\cellcolor[HTML]{E9E9E9} 1 &\cellcolor[HTML]{E9E9E9} 1 &\cellcolor[HTML]{E9E9E9} 4 &\cellcolor[HTML]{E9E9E9} 8 \\
3 &\cellcolor[HTML]{E9E9E9} 4 &\cellcolor[HTML]{E9E9E9} 4 &\cellcolor[HTML]{C7C5C5} 8 &\cellcolor[HTML]{C7C5C5} 4 \\
4 &\cellcolor[HTML]{E9E9E9} 8 &\cellcolor[HTML]{E9E9E9} 8 &\cellcolor[HTML]{E9E9E9} 6 &\cellcolor[HTML]{E9E9E9} 1 \\
5 &\cellcolor[HTML]{C7C5C5} 9 &\cellcolor[HTML]{C7C5C5} 9 &\cellcolor[HTML]{C7C5C5} 9 &\cellcolor[HTML]{C7C5C5} 9 \\
6 &\cellcolor[HTML]{E9E9E9} 2 &\cellcolor[HTML]{E9E9E9} 2 &\cellcolor[HTML]{E9E9E9} 7&\cellcolor[HTML]{E9E9E9} 3 \\
7 &\cellcolor[HTML]{E9E9E9} 3 &\cellcolor[HTML]{E9E9E9} 3 &\cellcolor[HTML]{C7C5C5} 2 &\cellcolor[HTML]{E9E9E9} 5 \\
8 &\cellcolor[HTML]{E9E9E9} 5 &\cellcolor[HTML]{E9E9E9} 5 &\cellcolor[HTML]{E9E9E9} 10 &\cellcolor[HTML]{C7C5C5} 7 \\
9 &\cellcolor[HTML]{E9E9E9} 7 &\cellcolor[HTML]{E9E9E9} 7 &\cellcolor[HTML]{C7C5C5} 3 &\cellcolor[HTML]{E9E9E9} 10 \\
10 &\cellcolor[HTML]{E9E9E9} 10 &\cellcolor[HTML]{E9E9E9} 10 &\cellcolor[HTML]{C7C5C5} 5 &\cellcolor[HTML]{C7C5C5} 2 \\\hline
\end{tabular}
    \begin{tablenotes}
      \small
      \item * nodes 4 and 8 slightly different; nodes 7 and 10 slightly different.
    \end{tablenotes}
\end{threeparttable}
\end{table}

\begin{figure}[ht!]
	\centering
\includegraphics[width=9cm]{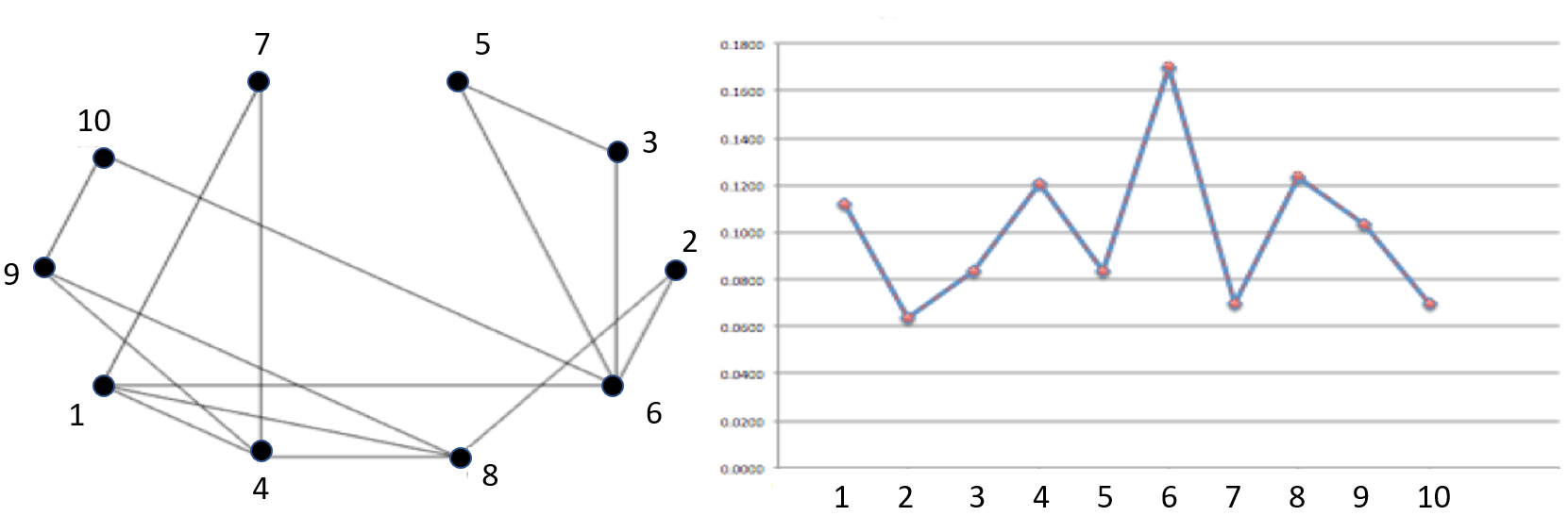}
\caption{Inconsistency Graph with ten nodes and energy distribution on each node after applying the Energy Flow method}
	\label{fig:EF2}
\end{figure}

\color{black}  
 
 Then we considered a more complicated situation as shown in Figure~\ref{fig:EF2}(left). It also shows the final distribution of energy per node (right), where again higher energy values correspond to nodes with lower reliability. The statement that inconsistency with less reliable nodes is less severe than inconsistency with more reliable nodes is also shown in this case. 
 
 For example, Node 2 has the same connectivity as Node 3, but is more reliable with less energy. The reason is that we made Node 2 inconsistent with Node 8 and also Node 3 inconsistent with Node 5 (which is more reliable than Node 8, since it has lower connectivity).
 
\color{black}  
 
 We also applied the two Max Voting methods and Hub Authority to calculate the rank of the nodes in this graph. The result is shown in Table~\ref{tab:compare3}. This time the two Max Voting methods give the same result. Compared to FaNDS, the Max Voting methods can not discriminate the group of (1, 4, 8) and (2, 3, 5, 7, 9, 10), which have the same number of edges but connect to different nodes. The rank from Hub Authority method is totally confusing in this case.

\section{Summary}

We have developed a new method, FaNDS, for detecting fake news, an issue that has
become significant in recent years.
We explained how FaNDS works using two major concepts: an Inconsistency Graph that is analyzed by the Energy Flow method.
We demonstrated our technique on the experimental data from the Fake News Challenge database, FNC-1.
We then showed that FaNDS is superior to several other fake news detection methods in its sensitivity and accuracy.


\label{sec:conclusion}

\section{Acknowledgments}
 This work is partially supported by NSF BCS-1244672 grant.

\bibliography{IP_frame_work}

\begin{thebibliography}{76}
\providecommand{\natexlab}[1]{#1}
\providecommand{\url}[1]{\texttt{#1}}
\providecommand{\href}[2]{#2}
\providecommand{\path}[1]{#1}
\providecommand{\DOIprefix}{doi:}
\providecommand{\ArXivprefix}{arXiv:}
\providecommand{\URLprefix}{URL: }
\providecommand{\Pubmedprefix}{pmid:}
\providecommand{\doi}[1]{\href{http://dx.doi.org/#1}{\path{#1}}}
\providecommand{\Pubmed}[1]{\href{pmid:#1}{\path{#1}}}
\providecommand{\BIBand}{and}
\providecommand{\bibinfo}[2]{#2}
\ifx\xfnm\undefined \def\xfnm[#1]{\unskip,\space#1}\fi
\makeatletter\def\@biblabel#1{#1.}\makeatother
\bibitem[{Pierri and Ceri(2019)}]{pierri2019false}
\bibinfo{author}{Pierri\xfnm[ F.]}, \bibinfo{author}{Ceri\xfnm[ S.]}.
\newblock \bibinfo{title}{False news on social media: A data-driven survey}.
\newblock \emph{\bibinfo{journal}{ACM SIGMOD Record}}
  \bibinfo{year}{2019};\bibinfo{volume}{48}(\bibinfo{number}{2}):\bibinfo{pages}{18--27}.
\bibitem[{Allcott and Gentzkow(2017)}]{allcott2017social}
\bibinfo{author}{Allcott\xfnm[ H.]}, \bibinfo{author}{Gentzkow\xfnm[ M.]}.
\newblock \bibinfo{title}{Social media and fake news in the 2016 election}.
\newblock \emph{\bibinfo{journal}{Journal of economic perspectives}}
  \bibinfo{year}{2017};\bibinfo{volume}{31}(\bibinfo{number}{2}):\bibinfo{pages}{211--36}.
\bibitem[{Shu et~al.(2019)Shu, Wang and Liu}]{shu2019beyond}
\bibinfo{author}{Shu\xfnm[ K.]}, \bibinfo{author}{Wang\xfnm[ S.]},
  \bibinfo{author}{Liu\xfnm[ H.]}.
\newblock \bibinfo{title}{Beyond news contents: The role of social context for
  fake news detection}.
\newblock In: \emph{\bibinfo{booktitle}{Proceedings of the Twelfth ACM
  International Conference on Web Search and Data Mining}}.
  \bibinfo{year}{2019}:\unskip \bibinfo{pages}{312--320}.
\bibitem[{Bondielli and Marcelloni(2019)}]{bondielli2019survey}
\bibinfo{author}{Bondielli\xfnm[ A.]}, \bibinfo{author}{Marcelloni\xfnm[ F.]}.
\newblock \bibinfo{title}{A survey on fake news and rumour detection
  techniques}.
\newblock \emph{\bibinfo{journal}{Information Sciences}}
  \bibinfo{year}{2019};\bibinfo{volume}{497}:\bibinfo{pages}{38--55}.
\bibitem[{DiFonzo and Bordia(2007)}]{difonzo2007rumor}
\bibinfo{author}{DiFonzo\xfnm[ N.]}, \bibinfo{author}{Bordia\xfnm[ P.]}.
\newblock \bibinfo{title}{Rumor, gossip and urban legends}.
\newblock \emph{\bibinfo{journal}{Diogenes}}
  \bibinfo{year}{2007};\bibinfo{volume}{54}(\bibinfo{number}{1}):\bibinfo{pages}{19--35}.
\bibitem[{Zubiaga et~al.(2015)Zubiaga, Liakata, Procter, Bontcheva and
  Tolmie}]{zubiaga2015towards}
\bibinfo{author}{Zubiaga\xfnm[ A.]}, \bibinfo{author}{Liakata\xfnm[ M.]},
  \bibinfo{author}{Procter\xfnm[ R.]}, \bibinfo{author}{Bontcheva\xfnm[ K.]},
  \bibinfo{author}{Tolmie\xfnm[ P.]}.
\newblock \bibinfo{title}{Towards detecting rumours in social media}.
\newblock In: \emph{\bibinfo{booktitle}{Workshops at the Twenty-Ninth AAAI
  Conference on Artificial Intelligence}}. \bibinfo{year}{2015}:\unskip.
\bibitem[{{The Museum of Hoaxes}(1997)}]{GreatMoonhoax}
\bibinfo{author}{{The Museum of Hoaxes}\xfnm[]}.
\newblock \bibinfo{title}{The great moon hoax}.
\newblock
  \bibinfo{howpublished}{\url{http://hoaxes.org/archive/permalink/the_great_moon_hoax}};
  \bibinfo{year}{1997}.
\bibitem[{Allport and Postman(1946)}]{Allport1946501}
\bibinfo{author}{Allport\xfnm[ G.]}, \bibinfo{author}{Postman\xfnm[ L.]}.
\newblock \bibinfo{title}{An analysis of rumor}.
\newblock \emph{\bibinfo{journal}{Public Opinion Quarterly}}
  \bibinfo{year}{1946};\bibinfo{volume}{10}(\bibinfo{number}{4}):\bibinfo{pages}{501--517}.
\newblock \URLprefix
  \url{https://www.scopus.com/inward/record.uri?eid=2-s2.0-0007285236&doi=10.1093%2fpoq%2f10.4.501&partnerID=40&md5=1d72343b350055e336bc67e36b115129}.
  \DOIprefix\doi{10.1093/poq/10.4.501}; \bibinfo{note}{cited By 87}.
\bibitem[{Allport and Postman(1947)}]{Allport1947240}
\bibinfo{author}{Allport\xfnm[ G.]}, \bibinfo{author}{Postman\xfnm[ L.]}.
\newblock \bibinfo{title}{The psychology of rumor}.
\newblock \emph{\bibinfo{journal}{ANNALS Am Acad Political Soc Sci}}
  \bibinfo{year}{1947};\bibinfo{volume}{257}(\bibinfo{number}{1}):\bibinfo{pages}{240--241}.
\newblock \bibinfo{note}{Cited By 1}.
\bibitem[{Zubiaga et~al.(2018{\natexlab{a}})Zubiaga, Aker, Bontcheva, Liakata
  and Procter}]{zubiaga2018detection}
\bibinfo{author}{Zubiaga\xfnm[ A.]}, \bibinfo{author}{Aker\xfnm[ A.]},
  \bibinfo{author}{Bontcheva\xfnm[ K.]}, \bibinfo{author}{Liakata\xfnm[ M.]},
  \bibinfo{author}{Procter\xfnm[ R.]}.
\newblock \bibinfo{title}{Detection and resolution of rumours in social media:
  A survey}.
\newblock \emph{\bibinfo{journal}{ACM Computing Surveys (CSUR)}}
  \bibinfo{year}{2018}{\natexlab{a}};\bibinfo{volume}{51}(\bibinfo{number}{2}):\bibinfo{pages}{1--36}.
\bibitem[{Diakopoulos et~al.(2012)Diakopoulos, De~Choudhury and
  Naaman}]{diakopoulos2012finding}
\bibinfo{author}{Diakopoulos\xfnm[ N.]}, \bibinfo{author}{De~Choudhury\xfnm[
  M.]}, \bibinfo{author}{Naaman\xfnm[ M.]}.
\newblock \bibinfo{title}{Finding and assessing social media information
  sources in the context of journalism}.
\newblock In: \emph{\bibinfo{booktitle}{Proceedings of the SIGCHI conference on
  human factors in computing systems}}. \bibinfo{year}{2012}:\unskip
  \bibinfo{pages}{2451--2460}.
\bibitem[{Tolmie et~al.(2017)Tolmie, Procter, Randall, Rouncefield, Burger,
  Wong Sak~Hoi, Zubiaga and Liakata}]{tolmie2017supporting}
\bibinfo{author}{Tolmie\xfnm[ P.]}, \bibinfo{author}{Procter\xfnm[ R.]},
  \bibinfo{author}{Randall\xfnm[ D.W.]}, \bibinfo{author}{Rouncefield\xfnm[
  M.]}, \bibinfo{author}{Burger\xfnm[ C.]}, \bibinfo{author}{Wong Sak~Hoi\xfnm[
  G.]}, \bibinfo{author}{Zubiaga\xfnm[ A.]}, \bibinfo{author}{Liakata\xfnm[
  M.]}.
\newblock \bibinfo{title}{Supporting the use of user generated content in
  journalistic practice}.
\newblock In: \emph{\bibinfo{booktitle}{Proceedings of the 2017 chi conference
  on human factors in computing systems}}. \bibinfo{year}{2017}:\unskip
  \bibinfo{pages}{3632--3644}.
\bibitem[{Hermida(2010)}]{hermida2010twittering}
\bibinfo{author}{Hermida\xfnm[ A.]}.
\newblock \bibinfo{title}{Twittering the news: The emergence of ambient
  journalism}.
\newblock \emph{\bibinfo{journal}{Journalism practice}}
  \bibinfo{year}{2010};\bibinfo{volume}{4}(\bibinfo{number}{3}):\bibinfo{pages}{297--308}.
\bibitem[{Vieweg(2010)}]{vieweg2010microblogged}
\bibinfo{author}{Vieweg\xfnm[ S.]}.
\newblock \bibinfo{title}{Microblogged contributions to the emergency arena:
  Discovery, interpretation and implications}.
\newblock \emph{\bibinfo{journal}{Computer Supported Collaborative Work}}
  \bibinfo{year}{2010};:\bibinfo{pages}{515--516}.
\bibitem[{Shao et~al.(2018)Shao, Ciampaglia, Varol, Yang, Flammini and
  Menczer}]{shao2018spread}
\bibinfo{author}{Shao\xfnm[ C.]}, \bibinfo{author}{Ciampaglia\xfnm[ G.L.]},
  \bibinfo{author}{Varol\xfnm[ O.]}, \bibinfo{author}{Yang\xfnm[ K.C.]},
  \bibinfo{author}{Flammini\xfnm[ A.]}, \bibinfo{author}{Menczer\xfnm[ F.]}.
\newblock \bibinfo{title}{The spread of low-credibility content by social
  bots}.
\newblock \emph{\bibinfo{journal}{Nature communications}}
  \bibinfo{year}{2018};\bibinfo{volume}{9}(\bibinfo{number}{1}):\bibinfo{pages}{1--9}.
\bibitem[{Kumar and Shah(2018)}]{kumar2018false}
\bibinfo{author}{Kumar\xfnm[ S.]}, \bibinfo{author}{Shah\xfnm[ N.]}.
\newblock \bibinfo{title}{False information on web and social media: A survey}.
\newblock \emph{\bibinfo{journal}{arXiv preprint arXiv:180408559}}
  \bibinfo{year}{2018};.
\bibitem[{Howard and Kollanyi(2016)}]{howard2016bots}
\bibinfo{author}{Howard\xfnm[ P.N.]}, \bibinfo{author}{Kollanyi\xfnm[ B.]}.
\newblock \bibinfo{title}{Bots,\# strongerin, and\# brexit: Computational
  propaganda during the uk-eu referendum}.
\newblock \emph{\bibinfo{journal}{arXiv preprint arXiv:160606356}}
  \bibinfo{year}{2016};.
\bibitem[{Ferrara(2017)}]{ferrara2017disinformation}
\bibinfo{author}{Ferrara\xfnm[ E.]}.
\newblock \bibinfo{title}{Disinformation and social bot operations in the run
  up to the 2017 french presidential election}.
\newblock \emph{\bibinfo{journal}{First Monday}}
  \bibinfo{year}{2017};\bibinfo{volume}{22}(\bibinfo{number}{8}).
\bibitem[{Matthews(2013)}]{matthews2013does}
\bibinfo{author}{Matthews\xfnm[ C.]}.
\newblock \bibinfo{title}{How does one fake tweet cause a stock market crash}.
\newblock \emph{\bibinfo{journal}{Wall Street \& Markets: Time}}
  \bibinfo{year}{2013};.
\bibitem[{Shu et~al.(2018)Shu, Mahudeswaran, Wang, Lee and
  Liu}]{shu2018fakenewsnet}
\bibinfo{author}{Shu\xfnm[ K.]}, \bibinfo{author}{Mahudeswaran\xfnm[ D.]},
  \bibinfo{author}{Wang\xfnm[ S.]}, \bibinfo{author}{Lee\xfnm[ D.]},
  \bibinfo{author}{Liu\xfnm[ H.]}.
\newblock \bibinfo{title}{Fakenewsnet: A data repository with news content,
  social context and spatialtemporal information for studying fake news on
  social media}.
\newblock \bibinfo{year}{2018}.
\newblock \href{http://arxiv.org/abs/1809.01286}{\tt arXiv:1809.01286}.
\bibitem[{Hauck(2017)}]{hauck2017pizzagate}
\bibinfo{author}{Hauck\xfnm[ G.]}.
\newblock \bibinfo{title}{'pizzagate' shooter sentenced to 4 years in prison}.
\newblock \bibinfo{year}{2017}.
\newblock \URLprefix
  \url{https://www.cnn.com/2017/06/22/politics/pizzagate-sentencing/index.html};
  \bibinfo{note}{last accessed 21 January 2020}.
\bibitem[{Ma et~al.(2016)Ma, Gao, Mitra, Kwon, Jansen, Wong and
  Cha}]{Ma20163818}
\bibinfo{author}{Ma\xfnm[ J.]}, \bibinfo{author}{Gao\xfnm[ W.]},
  \bibinfo{author}{Mitra\xfnm[ P.]}, \bibinfo{author}{Kwon\xfnm[ S.]},
  \bibinfo{author}{Jansen\xfnm[ B.]}, \bibinfo{author}{Wong\xfnm[ K.F.]},
  \bibinfo{author}{Cha\xfnm[ M.]}.
\newblock \bibinfo{title}{Detecting rumors from microblogs with recurrent
  neural networks}.
\newblock vol. \bibinfo{volume}{2016-January}. \bibinfo{year}{2016}:\unskip
  \bibinfo{pages}{3818--3824}.
\newblock \URLprefix
  \url{https://www.scopus.com/inward/record.uri?eid=2-s2.0-85006173435&partnerID=40&md5=bc9932e9d906fb5859df3efa69a13f7e};
  \bibinfo{note}{cited By 142}.
\bibitem[{Shu et~al.(2017)Shu, Sliva, Wang, Tang and Liu}]{shu2017fake}
\bibinfo{author}{Shu\xfnm[ K.]}, \bibinfo{author}{Sliva\xfnm[ A.]},
  \bibinfo{author}{Wang\xfnm[ S.]}, \bibinfo{author}{Tang\xfnm[ J.]},
  \bibinfo{author}{Liu\xfnm[ H.]}.
\newblock \bibinfo{title}{Fake news detection on social media: A data mining
  perspective}.
\newblock \emph{\bibinfo{journal}{ACM SIGKDD Explorations Newsletter}}
  \bibinfo{year}{2017};\bibinfo{volume}{19}(\bibinfo{number}{1}):\bibinfo{pages}{22--36}.
\bibitem[{Conroy et~al.(2015)Conroy, Rubin and Chen}]{conroy2015automatic}
\bibinfo{author}{Conroy\xfnm[ N.J.]}, \bibinfo{author}{Rubin\xfnm[ V.L.]},
  \bibinfo{author}{Chen\xfnm[ Y.]}.
\newblock \bibinfo{title}{Automatic deception detection: Methods for finding
  fake news}.
\newblock \emph{\bibinfo{journal}{Proceedings of the Association for
  Information Science and Technology}}
  \bibinfo{year}{2015};\bibinfo{volume}{52}(\bibinfo{number}{1}):\bibinfo{pages}{1--4}.
\bibitem[{Chen et~al.(2015)Chen, Conroy and Rubin}]{chen2015misleading}
\bibinfo{author}{Chen\xfnm[ Y.]}, \bibinfo{author}{Conroy\xfnm[ N.J.]},
  \bibinfo{author}{Rubin\xfnm[ V.L.]}.
\newblock \bibinfo{title}{Misleading online content: Recognizing clickbait as
  false news}.
\newblock In: \emph{\bibinfo{booktitle}{Proceedings of the 2015 ACM on Workshop
  on Multimodal Deception Detection}}. \bibinfo{organization}{ACM};
  \bibinfo{year}{2015}:\unskip \bibinfo{pages}{15--19}.
\bibitem[{Rubin and Lukoianova(2015)}]{rubin2015truth}
\bibinfo{author}{Rubin\xfnm[ V.L.]}, \bibinfo{author}{Lukoianova\xfnm[ T.]}.
\newblock \bibinfo{title}{Truth and deception at the rhetorical structure
  level}.
\newblock \emph{\bibinfo{journal}{Journal of the Association for Information
  Science and Technology}}
  \bibinfo{year}{2015};\bibinfo{volume}{66}(\bibinfo{number}{5}):\bibinfo{pages}{905--917}.
\bibitem[{Wang(2017)}]{wang2017liar}
\bibinfo{author}{Wang\xfnm[ W.Y.]}.
\newblock \bibinfo{title}{" liar, liar pants on fire": A new benchmark dataset
  for fake news detection}.
\newblock \emph{\bibinfo{journal}{arXiv preprint arXiv:170500648}}
  \bibinfo{year}{2017};.
\bibitem[{Hassan et~al.(2015)Hassan, Li and Tremayne}]{hassan2015detecting}
\bibinfo{author}{Hassan\xfnm[ N.]}, \bibinfo{author}{Li\xfnm[ C.]},
  \bibinfo{author}{Tremayne\xfnm[ M.]}.
\newblock \bibinfo{title}{Detecting check-worthy factual claims in presidential
  debates}.
\newblock In: \emph{\bibinfo{booktitle}{Proceedings of the 24th acm
  international on conference on information and knowledge management}}.
  \bibinfo{organization}{ACM}; \bibinfo{year}{2015}:\unskip
  \bibinfo{pages}{1835--1838}.
\bibitem[{Potthast et~al.(2017)Potthast, Kiesel, Reinartz, Bevendorff and
  Stein}]{potthast2017stylometric}
\bibinfo{author}{Potthast\xfnm[ M.]}, \bibinfo{author}{Kiesel\xfnm[ J.]},
  \bibinfo{author}{Reinartz\xfnm[ K.]}, \bibinfo{author}{Bevendorff\xfnm[ J.]},
  \bibinfo{author}{Stein\xfnm[ B.]}.
\newblock \bibinfo{title}{A stylometric inquiry into hyperpartisan and fake
  news}.
\newblock \emph{\bibinfo{journal}{arXiv preprint arXiv:170205638}}
  \bibinfo{year}{2017};.
\bibitem[{P{\'e}rez-Rosas et~al.(2017)P{\'e}rez-Rosas, Kleinberg, Lefevre and
  Mihalcea}]{perez2017automatic}
\bibinfo{author}{P{\'e}rez-Rosas\xfnm[ V.]}, \bibinfo{author}{Kleinberg\xfnm[
  B.]}, \bibinfo{author}{Lefevre\xfnm[ A.]}, \bibinfo{author}{Mihalcea\xfnm[
  R.]}.
\newblock \bibinfo{title}{Automatic detection of fake news}.
\newblock \emph{\bibinfo{journal}{arXiv preprint arXiv:170807104}}
  \bibinfo{year}{2017};.
\bibitem[{Ajao et~al.(2018)Ajao, Bhowmik and Zargari}]{ajao2018fake}
\bibinfo{author}{Ajao\xfnm[ O.]}, \bibinfo{author}{Bhowmik\xfnm[ D.]},
  \bibinfo{author}{Zargari\xfnm[ S.]}.
\newblock \bibinfo{title}{Fake news identification on twitter with hybrid cnn
  and rnn models}.
\newblock In: \emph{\bibinfo{booktitle}{Proceedings of the 9th International
  Conference on Social Media and Society}}. \bibinfo{year}{2018}:\unskip
  \bibinfo{pages}{226--230}.
\bibitem[{Kochkina et~al.(2018)Kochkina, Liakata and Zubiaga}]{kochkina2018all}
\bibinfo{author}{Kochkina\xfnm[ E.]}, \bibinfo{author}{Liakata\xfnm[ M.]},
  \bibinfo{author}{Zubiaga\xfnm[ A.]}.
\newblock \bibinfo{title}{All-in-one: Multi-task learning for rumour
  verification}.
\newblock \emph{\bibinfo{journal}{arXiv preprint arXiv:180603713}}
  \bibinfo{year}{2018};.
\bibitem[{Song et~al.(2019)Song, Yang, Chen, Tu, Liu and Sun}]{song2019ced}
\bibinfo{author}{Song\xfnm[ C.]}, \bibinfo{author}{Yang\xfnm[ C.]},
  \bibinfo{author}{Chen\xfnm[ H.]}, \bibinfo{author}{Tu\xfnm[ C.]},
  \bibinfo{author}{Liu\xfnm[ Z.]}, \bibinfo{author}{Sun\xfnm[ M.]}.
\newblock \bibinfo{title}{Ced: Credible early detection of social media
  rumors}.
\newblock \emph{\bibinfo{journal}{IEEE Transactions on Knowledge and Data
  Engineering}} \bibinfo{year}{2019};.
\bibitem[{Zubiaga et~al.(2018{\natexlab{b}})Zubiaga, Kochkina, Liakata,
  Procter, Lukasik, Bontcheva, Cohn and Augenstein}]{zubiaga2018discourse}
\bibinfo{author}{Zubiaga\xfnm[ A.]}, \bibinfo{author}{Kochkina\xfnm[ E.]},
  \bibinfo{author}{Liakata\xfnm[ M.]}, \bibinfo{author}{Procter\xfnm[ R.]},
  \bibinfo{author}{Lukasik\xfnm[ M.]}, \bibinfo{author}{Bontcheva\xfnm[ K.]},
  \bibinfo{author}{Cohn\xfnm[ T.]}, \bibinfo{author}{Augenstein\xfnm[ I.]}.
\newblock \bibinfo{title}{Discourse-aware rumour stance classification in
  social media using sequential classifiers}.
\newblock \emph{\bibinfo{journal}{Information Processing \& Management}}
  \bibinfo{year}{2018}{\natexlab{b}};\bibinfo{volume}{54}(\bibinfo{number}{2}):\bibinfo{pages}{273--290}.
\bibitem[{Castillo et~al.(2011)Castillo, Mendoza and
  Poblete}]{castillo2011information}
\bibinfo{author}{Castillo\xfnm[ C.]}, \bibinfo{author}{Mendoza\xfnm[ M.]},
  \bibinfo{author}{Poblete\xfnm[ B.]}.
\newblock \bibinfo{title}{Information credibility on twitter}.
\newblock In: \emph{\bibinfo{booktitle}{Proceedings of the 20th international
  conference on World wide web}}. \bibinfo{year}{2011}:\unskip
  \bibinfo{pages}{675--684}.
\bibitem[{Chu et~al.(2010)Chu, Gianvecchio, Wang and Jajodia}]{chu2010tweeting}
\bibinfo{author}{Chu\xfnm[ Z.]}, \bibinfo{author}{Gianvecchio\xfnm[ S.]},
  \bibinfo{author}{Wang\xfnm[ H.]}, \bibinfo{author}{Jajodia\xfnm[ S.]}.
\newblock \bibinfo{title}{Who is tweeting on twitter: human, bot, or cyborg?}
\newblock In: \emph{\bibinfo{booktitle}{Proceedings of the 26th annual computer
  security applications conference}}. \bibinfo{organization}{ACM};
  \bibinfo{year}{2010}:\unskip \bibinfo{pages}{21--30}.
\bibitem[{Qazvinian et~al.(2011)Qazvinian, Rosengren, Radev and
  Mei}]{qazvinian2011rumor}
\bibinfo{author}{Qazvinian\xfnm[ V.]}, \bibinfo{author}{Rosengren\xfnm[ E.]},
  \bibinfo{author}{Radev\xfnm[ D.R.]}, \bibinfo{author}{Mei\xfnm[ Q.]}.
\newblock \bibinfo{title}{Rumor has it: Identifying misinformation in
  microblogs}.
\newblock In: \emph{\bibinfo{booktitle}{Proceedings of the conference on
  empirical methods in natural language processing}}.
  \bibinfo{organization}{Association for Computational Linguistics};
  \bibinfo{year}{2011}:\unskip \bibinfo{pages}{1589--1599}.
\bibitem[{Kwon et~al.(2013)Kwon, Cha, Jung, Chen and Wang}]{kwon2013prominent}
\bibinfo{author}{Kwon\xfnm[ S.]}, \bibinfo{author}{Cha\xfnm[ M.]},
  \bibinfo{author}{Jung\xfnm[ K.]}, \bibinfo{author}{Chen\xfnm[ W.]},
  \bibinfo{author}{Wang\xfnm[ Y.]}.
\newblock \bibinfo{title}{Prominent features of rumor propagation in online
  social media}.
\newblock In: \emph{\bibinfo{booktitle}{2013 IEEE 13th International Conference
  on Data Mining}}. \bibinfo{organization}{IEEE}; \bibinfo{year}{2013}:\unskip
  \bibinfo{pages}{1103--1108}.
\bibitem[{Ma et~al.(2015)Ma, Gao, Wei, Lu and Wong}]{ma2015detect}
\bibinfo{author}{Ma\xfnm[ J.]}, \bibinfo{author}{Gao\xfnm[ W.]},
  \bibinfo{author}{Wei\xfnm[ Z.]}, \bibinfo{author}{Lu\xfnm[ Y.]},
  \bibinfo{author}{Wong\xfnm[ K.F.]}.
\newblock \bibinfo{title}{Detect rumors using time series of social context
  information on microblogging websites}.
\newblock In: \emph{\bibinfo{booktitle}{Proceedings of the 24th ACM
  International on Conference on Information and Knowledge Management}}.
  \bibinfo{year}{2015}:\unskip \bibinfo{pages}{1751--1754}.
\bibitem[{Kumar et~al.(2016)Kumar, West and Leskovec}]{kumar2016disinformation}
\bibinfo{author}{Kumar\xfnm[ S.]}, \bibinfo{author}{West\xfnm[ R.]},
  \bibinfo{author}{Leskovec\xfnm[ J.]}.
\newblock \bibinfo{title}{Disinformation on the web: Impact, characteristics,
  and detection of wikipedia hoaxes}.
\newblock In: \emph{\bibinfo{booktitle}{Proceedings of the 25th international
  conference on World Wide Web}}. \bibinfo{year}{2016}:\unskip
  \bibinfo{pages}{591--602}.
\bibitem[{Liu et~al.(2019)Liu, Jin and Shen}]{liu2019towards}
\bibinfo{author}{Liu\xfnm[ Y.]}, \bibinfo{author}{Jin\xfnm[ X.]},
  \bibinfo{author}{Shen\xfnm[ H.]}.
\newblock \bibinfo{title}{Towards early identification of online rumors based
  on long short-term memory networks}.
\newblock \emph{\bibinfo{journal}{Information Processing \& Management}}
  \bibinfo{year}{2019};\bibinfo{volume}{56}(\bibinfo{number}{4}):\bibinfo{pages}{1457--1467}.
\bibitem[{Li et~al.(2019)Li, Zhang and Si}]{li2019rumor}
\bibinfo{author}{Li\xfnm[ Q.]}, \bibinfo{author}{Zhang\xfnm[ Q.]},
  \bibinfo{author}{Si\xfnm[ L.]}.
\newblock \bibinfo{title}{Rumor detection by exploiting user credibility
  information, attention and multi-task learning}.
\newblock In: \emph{\bibinfo{booktitle}{Proceedings of the 57th Annual Meeting
  of the Association for Computational Linguistics}}.
  \bibinfo{year}{2019}:\unskip \bibinfo{pages}{1173--1179}.
\bibitem[{Gupta et~al.(2012)Gupta, Zhao and Han}]{gupta2012evaluating}
\bibinfo{author}{Gupta\xfnm[ M.]}, \bibinfo{author}{Zhao\xfnm[ P.]},
  \bibinfo{author}{Han\xfnm[ J.]}.
\newblock \bibinfo{title}{Evaluating event credibility on twitter}.
\newblock In: \emph{\bibinfo{booktitle}{Proceedings of the 2012 SIAM
  International Conference on Data Mining}}. \bibinfo{organization}{SIAM};
  \bibinfo{year}{2012}:\unskip \bibinfo{pages}{153--164}.
\bibitem[{Jin et~al.(2016)Jin, Cao, Zhang and Luo}]{jin2016news}
\bibinfo{author}{Jin\xfnm[ Z.]}, \bibinfo{author}{Cao\xfnm[ J.]},
  \bibinfo{author}{Zhang\xfnm[ Y.]}, \bibinfo{author}{Luo\xfnm[ J.]}.
\newblock \bibinfo{title}{News verification by exploiting conflicting social
  viewpoints in microblogs}.
\newblock In: \emph{\bibinfo{booktitle}{Thirtieth AAAI conference on artificial
  intelligence}}. \bibinfo{year}{2016}:\unskip.
\bibitem[{Ruchansky et~al.(2017)Ruchansky, Seo and Liu}]{ruchansky2017csi}
\bibinfo{author}{Ruchansky\xfnm[ N.]}, \bibinfo{author}{Seo\xfnm[ S.]},
  \bibinfo{author}{Liu\xfnm[ Y.]}.
\newblock \bibinfo{title}{Csi: A hybrid deep model for fake news detection}.
\newblock In: \emph{\bibinfo{booktitle}{Proceedings of the 2017 ACM on
  Conference on Information and Knowledge Management}}.
  \bibinfo{year}{2017}:\unskip \bibinfo{pages}{797--806}.
\bibitem[{Tacchini et~al.(2017)Tacchini, Ballarin, Della~Vedova, Moret and
  de~Alfaro}]{tacchini2017some}
\bibinfo{author}{Tacchini\xfnm[ E.]}, \bibinfo{author}{Ballarin\xfnm[ G.]},
  \bibinfo{author}{Della~Vedova\xfnm[ M.L.]}, \bibinfo{author}{Moret\xfnm[
  S.]}, \bibinfo{author}{de~Alfaro\xfnm[ L.]}.
\newblock \bibinfo{title}{Some like it hoax: Automated fake news detection in
  social networks}.
\newblock \emph{\bibinfo{journal}{arXiv preprint arXiv:170407506}}
  \bibinfo{year}{2017};.
\bibitem[{Della~Vedova et~al.(2018)Della~Vedova, Tacchini, Moret, Ballarin,
  DiPierro and de~Alfaro}]{della2018automatic}
\bibinfo{author}{Della~Vedova\xfnm[ M.L.]}, \bibinfo{author}{Tacchini\xfnm[
  E.]}, \bibinfo{author}{Moret\xfnm[ S.]}, \bibinfo{author}{Ballarin\xfnm[
  G.]}, \bibinfo{author}{DiPierro\xfnm[ M.]}, \bibinfo{author}{de~Alfaro\xfnm[
  L.]}.
\newblock \bibinfo{title}{Automatic online fake news detection combining
  content and social signals}.
\newblock In: \emph{\bibinfo{booktitle}{2018 22nd Conference of Open
  Innovations Association (FRUCT)}}. \bibinfo{organization}{IEEE};
  \bibinfo{year}{2018}:\unskip \bibinfo{pages}{272--279}.
\bibitem[{Guacho et~al.(2018)Guacho, Abdali, Shah and
  Papalexakis}]{guacho2018semi}
\bibinfo{author}{Guacho\xfnm[ G.B.]}, \bibinfo{author}{Abdali\xfnm[ S.]},
  \bibinfo{author}{Shah\xfnm[ N.]}, \bibinfo{author}{Papalexakis\xfnm[ E.E.]}.
\newblock \bibinfo{title}{Semi-supervised content-based detection of
  misinformation via tensor embeddings}.
\newblock In: \emph{\bibinfo{booktitle}{2018 IEEE/ACM International Conference
  on Advances in Social Networks Analysis and Mining (ASONAM)}}.
  \bibinfo{organization}{IEEE}; \bibinfo{year}{2018}:\unskip
  \bibinfo{pages}{322--325}.
\bibitem[{Wu et~al.(2014)Wu, Agarwal, Li, Yang and Yu}]{wu2014toward}
\bibinfo{author}{Wu\xfnm[ Y.]}, \bibinfo{author}{Agarwal\xfnm[ P.K.]},
  \bibinfo{author}{Li\xfnm[ C.]}, \bibinfo{author}{Yang\xfnm[ J.]},
  \bibinfo{author}{Yu\xfnm[ C.]}.
\newblock \bibinfo{title}{Toward computational fact-checking}.
\newblock \emph{\bibinfo{journal}{Proceedings of the VLDB Endowment}}
  \bibinfo{year}{2014};\bibinfo{volume}{7}(\bibinfo{number}{7}):\bibinfo{pages}{589--600}.
\bibitem[{Ciampaglia et~al.(2015)Ciampaglia, Shiralkar, Rocha, Bollen, Menczer
  and Flammini}]{ciampaglia2015computational}
\bibinfo{author}{Ciampaglia\xfnm[ G.L.]}, \bibinfo{author}{Shiralkar\xfnm[
  P.]}, \bibinfo{author}{Rocha\xfnm[ L.M.]}, \bibinfo{author}{Bollen\xfnm[
  J.]}, \bibinfo{author}{Menczer\xfnm[ F.]}, \bibinfo{author}{Flammini\xfnm[
  A.]}.
\newblock \bibinfo{title}{Computational fact checking from knowledge networks}.
\newblock \emph{\bibinfo{journal}{PloS one}}
  \bibinfo{year}{2015};\bibinfo{volume}{10}(\bibinfo{number}{6}):\bibinfo{pages}{e0128193}.
\bibitem[{Shi and Weninger(2016)}]{shi2016fact}
\bibinfo{author}{Shi\xfnm[ B.]}, \bibinfo{author}{Weninger\xfnm[ T.]}.
\newblock \bibinfo{title}{Fact checking in heterogeneous information networks}.
\newblock In: \emph{\bibinfo{booktitle}{Proceedings of the 25th International
  Conference Companion on World Wide Web}}.
  \bibinfo{organization}{International World Wide Web Conferences Steering
  Committee}; \bibinfo{year}{2016}:\unskip \bibinfo{pages}{101--102}.
\bibitem[{Jin et~al.(2014)Jin, Cao, Jiang and Zhang}]{jin2014news}
\bibinfo{author}{Jin\xfnm[ Z.]}, \bibinfo{author}{Cao\xfnm[ J.]},
  \bibinfo{author}{Jiang\xfnm[ Y.G.]}, \bibinfo{author}{Zhang\xfnm[ Y.]}.
\newblock \bibinfo{title}{News credibility evaluation on microblog with a
  hierarchical propagation model}.
\newblock In: \emph{\bibinfo{booktitle}{2014 IEEE International Conference on
  Data Mining}}. \bibinfo{organization}{IEEE}; \bibinfo{year}{2014}:\unskip
  \bibinfo{pages}{230--239}.
\bibitem[{Mukherjee and Weikum(2015)}]{mukherjee2015leveraging}
\bibinfo{author}{Mukherjee\xfnm[ S.]}, \bibinfo{author}{Weikum\xfnm[ G.]}.
\newblock \bibinfo{title}{Leveraging joint interactions for credibility
  analysis in news communities}.
\newblock In: \emph{\bibinfo{booktitle}{Proceedings of the 24th ACM
  International on Conference on Information and Knowledge Management}}.
  \bibinfo{organization}{ACM}; \bibinfo{year}{2015}:\unskip
  \bibinfo{pages}{353--362}.
\bibitem[{J{\o}sang et~al.(2007)J{\o}sang, Ismail and Boyd}]{josang2007survey}
\bibinfo{author}{J{\o}sang\xfnm[ A.]}, \bibinfo{author}{Ismail\xfnm[ R.]},
  \bibinfo{author}{Boyd\xfnm[ C.]}.
\newblock \bibinfo{title}{A survey of trust and reputation systems for online
  service provision}.
\newblock \emph{\bibinfo{journal}{Decision support systems}}
  \bibinfo{year}{2007};\bibinfo{volume}{43}(\bibinfo{number}{2}):\bibinfo{pages}{618--644}.
\bibitem[{Page et~al.(1999)Page, Brin, Motwani and Winograd}]{page1999pagerank}
\bibinfo{author}{Page\xfnm[ L.]}, \bibinfo{author}{Brin\xfnm[ S.]},
  \bibinfo{author}{Motwani\xfnm[ R.]}, \bibinfo{author}{Winograd\xfnm[ T.]}.
\newblock \bibinfo{title}{The pagerank citation ranking: Bringing order to the
  web.}
\newblock \bibinfo{type}{Tech. Rep.}; Stanford InfoLab; \bibinfo{year}{1999}.
\bibitem[{Kleinberg(1999)}]{kleinberg1999hubs}
\bibinfo{author}{Kleinberg\xfnm[ J.M.]}.
\newblock \bibinfo{title}{Hubs, authorities, and communities}.
\newblock \emph{\bibinfo{journal}{ACM computing surveys (CSUR)}}
  \bibinfo{year}{1999};\bibinfo{volume}{31}(\bibinfo{number}{4es}):\bibinfo{pages}{5}.
\bibitem[{Lempel and Moran(2000)}]{lempel2000stochastic}
\bibinfo{author}{Lempel\xfnm[ R.]}, \bibinfo{author}{Moran\xfnm[ S.]}.
\newblock \bibinfo{title}{The stochastic approach for link-structure analysis
  (salsa) and the tkc effect}.
\newblock \emph{\bibinfo{journal}{Computer Networks}}
  \bibinfo{year}{2000};\bibinfo{volume}{33}(\bibinfo{number}{1-6}):\bibinfo{pages}{387--401}.
\bibitem[{Ceglowski et~al.(2003)Ceglowski, Coburn and
  Cuadrado}]{ceglowski2003semantic}
\bibinfo{author}{Ceglowski\xfnm[ M.]}, \bibinfo{author}{Coburn\xfnm[ A.]},
  \bibinfo{author}{Cuadrado\xfnm[ J.]}.
\newblock \bibinfo{title}{Semantic search of unstructured data using contextual
  network graphs}.
\newblock \emph{\bibinfo{journal}{National Institute for Technology and Liberal
  Education}} \bibinfo{year}{2003};\bibinfo{volume}{10}.
\bibitem[{Levien(2004)}]{levien2004attack}
\bibinfo{author}{Levien\xfnm[ R.]}.
\newblock \bibinfo{title}{Attack-resistant trust metrics. ph. d. thesis}.
\newblock \emph{\bibinfo{journal}{University of California at Berkeley, USA, Ph
  D thesis}} \bibinfo{year}{2004};.
\bibitem[{Ziegler and Lausen(2004)}]{ziegler2004spreading}
\bibinfo{author}{Ziegler\xfnm[ C.N.]}, \bibinfo{author}{Lausen\xfnm[ G.]}.
\newblock \bibinfo{title}{Spreading activation models for trust propagation}.
\newblock In: \emph{\bibinfo{booktitle}{IEEE International Conference on
  e-Technology, e-Commerce and e-Service, 2004. EEE'04. 2004}}.
  \bibinfo{organization}{IEEE}; \bibinfo{year}{2004}:\unskip
  \bibinfo{pages}{83--97}.
\bibitem[{Ziegler and Lausen(2005)}]{ziegler2005propagation}
\bibinfo{author}{Ziegler\xfnm[ C.N.]}, \bibinfo{author}{Lausen\xfnm[ G.]}.
\newblock \bibinfo{title}{Propagation models for trust and distrust in social
  networks}.
\newblock \emph{\bibinfo{journal}{Information Systems Frontiers}}
  \bibinfo{year}{2005};\bibinfo{volume}{7}(\bibinfo{number}{4-5}):\bibinfo{pages}{337--358}.
\bibitem[{Kamvar et~al.(2003)Kamvar, Schlosser and
  Garcia-Molina}]{kamvar2003eigentrust}
\bibinfo{author}{Kamvar\xfnm[ S.D.]}, \bibinfo{author}{Schlosser\xfnm[ M.T.]},
  \bibinfo{author}{Garcia-Molina\xfnm[ H.]}.
\newblock \bibinfo{title}{The eigentrust algorithm for reputation management in
  p2p networks}.
\newblock In: \emph{\bibinfo{booktitle}{Proceedings of the 12th international
  conference on World Wide Web}}. \bibinfo{organization}{ACM};
  \bibinfo{year}{2003}:\unskip \bibinfo{pages}{640--651}.
\bibitem[{Zhang et~al.(2019)Zhang, Zadorozhny and Oleshchuk}]{zhang2019slftd}
\bibinfo{author}{Zhang\xfnm[ D.]}, \bibinfo{author}{Zadorozhny\xfnm[ V.I.]},
  \bibinfo{author}{Oleshchuk\xfnm[ V.A.]}.
\newblock \bibinfo{title}{Slftd: A subjective logic based framework for truth
  discovery}.
\newblock In: \emph{\bibinfo{booktitle}{European Conference on Advances in
  Databases and Information Systems}}. \bibinfo{organization}{Springer};
  \bibinfo{year}{2019}:\unskip \bibinfo{pages}{102--110}.
\bibitem[{Gyongyi et~al.(2006)Gyongyi, Berkhin, Garcia-Molina and
  Pedersen}]{gyongyi2006link}
\bibinfo{author}{Gyongyi\xfnm[ Z.]}, \bibinfo{author}{Berkhin\xfnm[ P.]},
  \bibinfo{author}{Garcia-Molina\xfnm[ H.]}, \bibinfo{author}{Pedersen\xfnm[
  J.]}.
\newblock \bibinfo{title}{Link spam detection based on mass estimation}.
\newblock In: \emph{\bibinfo{booktitle}{Proceedings of the 32nd international
  conference on Very large data bases}}. \bibinfo{organization}{VLDB
  Endowment}; \bibinfo{year}{2006}:\unskip \bibinfo{pages}{439--450}.
\bibitem[{Xi et~al.(2004)Xi, Zhang, Chen, Lu, Yan, Ma and Fox}]{xi2004link}
\bibinfo{author}{Xi\xfnm[ W.]}, \bibinfo{author}{Zhang\xfnm[ B.]},
  \bibinfo{author}{Chen\xfnm[ Z.]}, \bibinfo{author}{Lu\xfnm[ Y.]},
  \bibinfo{author}{Yan\xfnm[ S.]}, \bibinfo{author}{Ma\xfnm[ W.Y.]},
  \bibinfo{author}{Fox\xfnm[ E.A.]}.
\newblock \bibinfo{title}{Link fusion: a unified link analysis framework for
  multi-type interrelated data objects}.
\newblock In: \emph{\bibinfo{booktitle}{Proceedings of the 13th international
  conference on World Wide Web}}. \bibinfo{organization}{ACM};
  \bibinfo{year}{2004}:\unskip \bibinfo{pages}{319--327}.
\bibitem[{Del~Corso et~al.(2005)Del~Corso, Gulli and Romani}]{del2005fast}
\bibinfo{author}{Del~Corso\xfnm[ G.M.]}, \bibinfo{author}{Gulli\xfnm[ A.]},
  \bibinfo{author}{Romani\xfnm[ F.]}.
\newblock \bibinfo{title}{Fast pagerank computation via a sparse linear
  system}.
\newblock \emph{\bibinfo{journal}{Internet Mathematics}}
  \bibinfo{year}{2005};\bibinfo{volume}{2}(\bibinfo{number}{3}):\bibinfo{pages}{251--273}.
\bibitem[{Berkhin(2005)}]{berkhin2005survey}
\bibinfo{author}{Berkhin\xfnm[ P.]}.
\newblock \bibinfo{title}{A survey on pagerank computing}.
\newblock \emph{\bibinfo{journal}{Internet Mathematics}}
  \bibinfo{year}{2005};\bibinfo{volume}{2}(\bibinfo{number}{1}):\bibinfo{pages}{73--120}.
\bibitem[{Xu et~al.(2019)Xu, Zadorozhny and Grant}]{xu2019incompfuse}
\bibinfo{author}{Xu\xfnm[ J.]}, \bibinfo{author}{Zadorozhny\xfnm[ V.]},
  \bibinfo{author}{Grant\xfnm[ J.]}.
\newblock \bibinfo{title}{Incompfuse: a logical framework for historical
  information fusion with inaccurate data sources}.
\newblock \emph{\bibinfo{journal}{Journal of Intelligent Information Systems}}
  \bibinfo{year}{2019};\DOIprefix\doi{https://doi.org/10.1016/j.is.2020.101508}.
\bibitem[{Xu et~al.(2020)Xu, Zadorozhny and Grant}]{XU2020101508}
\bibinfo{author}{Xu\xfnm[ J.]}, \bibinfo{author}{Zadorozhny\xfnm[ V.]},
  \bibinfo{author}{Grant\xfnm[ J.]}.
\newblock \bibinfo{title}{A-cure: An accurate information reconstruction from
  inaccurate data sources}.
\newblock \emph{\bibinfo{journal}{Information Systems}}
  \bibinfo{year}{2020};:\bibinfo{pages}{101508}\DOIprefix\doi{https://doi.org/10.1016/j.is.2020.101508}.
\bibitem[{Bostock et~al.(2011)Bostock, Ogievetsky and Heer}]{bostock2011d3}
\bibinfo{author}{Bostock\xfnm[ M.]}, \bibinfo{author}{Ogievetsky\xfnm[ V.]},
  \bibinfo{author}{Heer\xfnm[ J.]}.
\newblock \bibinfo{title}{D$^3$ data-driven documents}.
\newblock \emph{\bibinfo{journal}{IEEE transactions on visualization and
  computer graphics}}
  \bibinfo{year}{2011};\bibinfo{volume}{17}(\bibinfo{number}{12}):\bibinfo{pages}{2301--2309}.
\bibitem[{Bountouridis et~al.(2018)Bountouridis, Marrero, Tintarev and
  Hauff}]{bountouridis2018explaining}
\bibinfo{author}{Bountouridis\xfnm[ D.]}, \bibinfo{author}{Marrero\xfnm[ M.]},
  \bibinfo{author}{Tintarev\xfnm[ N.]}, \bibinfo{author}{Hauff\xfnm[ C.]}.
\newblock \bibinfo{title}{Explaining credibility in news articles using
  cross-referencing}.
\newblock In: \emph{\bibinfo{booktitle}{SIGIR workshop on ExplainAble
  Recommendation and Search (EARS)}}. \bibinfo{year}{2018}:\unskip.
\bibitem[{Baly et~al.(2018)Baly, Karadzhov, Alexandrov, Glass and
  Nakov}]{baly2018predicting}
\bibinfo{author}{Baly\xfnm[ R.]}, \bibinfo{author}{Karadzhov\xfnm[ G.]},
  \bibinfo{author}{Alexandrov\xfnm[ D.]}, \bibinfo{author}{Glass\xfnm[ J.]},
  \bibinfo{author}{Nakov\xfnm[ P.]}.
\newblock \bibinfo{title}{Predicting factuality of reporting and bias of news
  media sources}.
\newblock \emph{\bibinfo{journal}{arXiv preprint arXiv:181001765}}
  \bibinfo{year}{2018};.
\bibitem[{Fairbanks et~al.(2018)Fairbanks, Fitch, Knauf and
  Briscoe}]{fairbanks2018credibility}
\bibinfo{author}{Fairbanks\xfnm[ J.]}, \bibinfo{author}{Fitch\xfnm[ N.]},
  \bibinfo{author}{Knauf\xfnm[ N.]}, \bibinfo{author}{Briscoe\xfnm[ E.]}.
\newblock \bibinfo{title}{Credibility assessment in the news: Do we need to
  read}.
\newblock In: \emph{\bibinfo{booktitle}{Proc. of the MIS2 Workshop held in
  conjuction with 11th Int’l Conf. on Web Search and Data Mining}}.
  \bibinfo{year}{2018}:\unskip \bibinfo{pages}{799--800}.
\bibitem[{Heath(2016)}]{heath2016facebook}
\bibinfo{author}{Heath\xfnm[ A.]}.
\newblock \bibinfo{title}{Facebook is going to use snopes and other
  fact-checkers to combat and bury'fake news.'}.
\newblock \emph{\bibinfo{journal}{Business Insider}} \bibinfo{year}{2016};.
\bibitem[{Berghel(2017)}]{berghel2017lies}
\bibinfo{author}{Berghel\xfnm[ H.]}.
\newblock \bibinfo{title}{Lies, damn lies, and fake news}.
\newblock \emph{\bibinfo{journal}{Computer}}
  \bibinfo{year}{2017};\bibinfo{volume}{50}(\bibinfo{number}{2}):\bibinfo{pages}{80--85}.
\bibitem[{Bounegru et~al.(2018)Bounegru, Gray, Venturini and
  Mauri}]{bounegru2018field}
\bibinfo{author}{Bounegru\xfnm[ L.]}, \bibinfo{author}{Gray\xfnm[ J.]},
  \bibinfo{author}{Venturini\xfnm[ T.]}, \bibinfo{author}{Mauri\xfnm[ M.]}.
\newblock \bibinfo{title}{A field guide to'fake news' and other information
  disorders}.
\newblock \emph{\bibinfo{journal}{A Field Guide to" Fake News" and Other
  Information Disorders: A Collection of Recipes for Those Who Love to Cook
  with Digital Methods, Public Data Lab, Amsterdam (2018)}}
  \bibinfo{year}{2018};.

\end{thebibliography}

\end{document}